\documentstyle[12pt,equation]{article}
\advance\voffset by 28 pt
\begin{document}
\newcommand{\nav}{\mbox{$\langle n_{ch} \rangle$}}
\newcommand{\ptav}{\mbox{$\langle p_{T,ch} \rangle$}}
\newcommand{\ph}{\mbox{$P_{\rm had}$}}
\newcommand{\ppsoft}{\mbox{$\sigma_{p \bar p}^{\rm soft}$}}
\newcommand{\pphard}{\mbox{$\sigma_{p \bar p}^{\rm hard}$}}
\newcommand{\ggsoft}{\mbox{$\sigma_{\gamma \gamma}^{\rm soft}$}}
\newcommand{\gghard}{\mbox{$\sigma_{\gamma \gamma}^{\rm hard}$}}
\newcommand{\csoft}{\mbox{$\chi_{p \bar p}^{\rm soft}$}}
\newcommand{\ccbar}{\mbox{$c \overline{c} $}}
\newcommand{\bbbar}{\mbox{$b \overline{b} $}}
\newcommand{\ttbar}{\mbox{$t \overline{t} $}}
\newcommand{\Y}{\mbox{$\Upsilon$}}
\newcommand{\aem}{\mbox{$\alpha_{{\rm em}}$}}
\newcommand{\shat}{\mbox{$\hat{s}$}}
\newcommand{\lhat}{\mbox{$\hat{\cal L}$}}
\newcommand{\sighat}{\mbox{$\hat{\sigma}$}}
\newcommand{\rs}{\mbox{$\sqrt{s}$}}
\newcommand{\pT}{\mbox{$p_T$}}
\newcommand{\wgg}{\mbox{$W_{\gamma \gamma}$}}
\newcommand{\fge}{\mbox{$f_{\gamma|e}$}}
\newcommand{\fee}{\mbox{$f_{e|e}$}}
\newcommand{\ptmin}{\mbox{$p_{T,min}$}}
\newcommand{\qvec}{\mbox{$\vec{q}^{\gamma}$}}
\newcommand{\xq}{\mbox{$(x,Q^2)$}}
\newcommand{\qig}{\mbox{$q_i^{\gamma}$}}
\newcommand{\Gg}{\mbox{$G^{\gamma}$}}
\newcommand{\qqbar}{\mbox{$q \overline{q}$}}
\newcommand{\ppbar}{\mbox{$p \overline{p}$}}
\newcommand{\ee}{\mbox{$e^+e^-$}}
\newcommand{\gaga}{\mbox{$\gamma\gamma$}}
\newcommand{\be}{\begin{equation}}
\newcommand{\ene}{\end{equation}}
\newcommand{\een}{\end{subequations}}
\newcommand{\ben}{\begin{subequations}}
\newcommand{\beq}{\begin{eqalignno}}
\newcommand{\eeq}{\end{eqalignno}}
\renewcommand{\thefootnote}{\fnsymbol{footnote} }

\pagestyle{empty}
\noindent
DESY 92-044 \\
BU 92/1 \\
March 1992
\vspace{2cm}
\begin{center}
{\Large \bf Aspects of Two--Photon Physics at Linear \ee\ Colliders}\\
\vspace{5mm}
Manuel Drees\\%footnote{Donations to: Commerzbank Hamburg, BLZ 20040000, acct.
%no. 5091368}\\
{\em Deutsches Elektronen-Synchrotron DESY, W2000 Hamburg 52, Germany} \\
\vspace{5mm}
Rohini M. Godbole\\
{\em Dept. of Physics, Univ. of Bombay, Vidyanagari, Bombay 400098, India}
\end{center}

\vspace{2.cm}
\begin{abstract}
We discuss various reactions at future \ee\ and \gaga\ colliders
involving real (beamstrahlung or backscattered laser) or quasi--real
(bremsstrahlung) photons in the initial state and hadrons in the
final state. The production of two central jets with large transverse
momentum \pT\ is described in some detail; we give distributions for the
rapidity and \pT\ of the jets as well as the di--jet invariant mass, and
discuss the relative importance of various initial state configurations and
the uncertainties that arise from the at present rather poor knowledge of
the parton content of the photon. We also present results for `mono--jet'
production where one jet goes down a beam pipe, for the production of
charm, bottom and top quarks, and for single production of $W$ and $Z$
bosons. Where appropriate, the two--photon processes are compared with
annihilation reactions leading to similar final states. We also argue that
the behaviour of the total inelastic \gaga\ cross section at high energies
will probably have little impact on the severity of background problems
caused by soft and semi--hard (`minijet') two--photon reactions. We find
very large differences in cross sections for all two--photon processes
between existing designs for future \ee\ colliders, due
to the different beamstrahlung spectra; in particular, both designs with
$\ll 1$ and $\gg 1$ events per bunch crossing exist. The number of
hadronic two--photon events is expected to rise quickly with the beam energy.
Hadronic backgrounds will
be even worse if the \ee\ collider is converted into a \gaga\ collider.
\end{abstract}
\clearpage
\setcounter{page}{1}
\pagestyle{plain}
\section*{1. Introduction}
In recent years an increasing amount of effort has been devoted \cite{1,1a}
to the study
of the physics potential and design problems of future \ee\ colliders. There
is great physics interest in such machines, since it is quite likely that
the planned $pp$ supercolliders \cite{2} will not be very effective for
searches for many hypothetical new particles (supersymmetric sleptons and
gauginos; non--standard Higgs bosons; heavy leptons; \dots) which interact
only weakly and/or lack the distinct signatures necessary for their discovery
at hadron colliders. Moreover, now that the construction of TRISTAN, the first
phase of LEP and the SLC has been completed, the time is ripe to develop
specific plans for the next generation of \ee\ colliders.

Traditionally \ee\ colliders have offered a very clean experimental
environment, allowing for the detailed study of particles that may have
been discovered previously at a hadron collider; a recent example is the
$Z$ boson, which is now being studied in great detail at the SLC and LEP.
This is the second major physics motivation for pushing the energy
frontier of \ee\ colliders to higher values.

However, as the collision energy \rs\ is increased, the cross
section for the annihilation events whose study will be the main purpose of
any future collider decreases like $1/s$, or at best like $\log s/s$.
At the same time, the cross section for the simplest hard two--photon process,
$\ee \rightarrow \ee \qqbar$, {\em increases} like $\log^3 s$, for fixed
transverse momentum of the quarks or fixed invariant mass of the \qqbar\
pair. Moreover, the hadronic structure of the photon also plays an increasingly
important role at higher energies. It can be described by introducing \cite{3}
quark and gluon densities ``inside'' the photon. These give rise \cite{4,5}
to processes where the partons ``in'' the photon, rather than the photons
themselves, undergo hard scattering. The cross section for these ``resolved
photon'' processes are predicted \cite{6} to rise almost linearly with the
\ee\ cms energy. This rapid increase has recently been confirmed by the AMY
group \cite{7} in the PETRA to TRISTAN energy range, 30  GeV $\leq \rs
\leq 60$ GeV. These considerations imply that at future \ee\ colliders,
hard two--photon events will outnumber annihilation events by an increasingly
wide margin.

The number of two--photon events is further boosted by beamstrahlung \cite{8},
which could increase \cite{9} the two--photon luminosity by as much as a factor
of 100 already at $\rs = 500$ GeV. As well known, synchrotron radiation
makes it prohibitively expensive to build \ee\ storage rings with energies
significantly beyond that of the second stage of LEP, $\rs \simeq 200$ GeV.
At linear colliders any given bunch of electrons or positrons crosses the
interaction point only once, as compared to approximately $10^8$ times at
LEP; moreover, the luminosity has to increase proportional to $s$ in order to
maintain a constant rate of annihilation events. These considerations imply
that a very high luminosity per bunch crossing has to be achieved at \ee\
linacs. This forces one to use small, dense bunches; the particles in
each bunch are therefore subject to strong electromagnetic fields just
before and during the bunch collisions. The resulting forces on the particles
in the bunches lead to their rapid acceleration; beamstrahlung is the
radiation emitted by the accelerating electrons and positrons.

In a recent Letter \cite{10} we pointed out that the combination of the
rapid increase of the cross section for resolved photon processes and
the enhanced photon flux due to beamstrahlung can lead to  severe
hadronic backgrounds at \ee\ supercolliders. We demonstrated this using the
design of ref.\cite{11} for a collider operating at $\rs = 1$ TeV, which
is characterized by a hard beamstrahlung spectrum. Under this assumption,
two--photon events dominate total dijet production for $\pT \leq 200$ GeV.
What is worse, one might have to expect ${\cal O}(10)$ semi--hard two--photon
events at {\em each} bunch crossing; this would give rise to an
``underlying event'' depositing as much as 100 GeV of transverse energy in the
detector.

Since then, more realistic designs for \ee\ colliders operating at \rs\ =
500 GeV have been put forward \cite{1a}. Unlike the example of ref.\cite{11},
these designs all foresee flat beams; furthermore, they split the bunch
into a bunch train consisting of several micro--bunches, which reduces the
necessary luminosity per bunch crossing. Both modifications reduce
beamstrahlung. In this paper we study representative examples of these recent
designs. We find very large differences between them, as far as the severity
of two--photon induced backgrounds are concerned. While for one proposed
design the situation is almost as problematic as for the ``theoretical''
collider of ref.\cite{11}, other designs, most noteably one using
superconducting cavities, are almost free of beamstrahlung--induced
hadronic backgrounds.

It has been suggested \cite{12} to
convert future \ee\ colliders into \gaga\ colliders. This can be
achieved by bathing the incoming $e^+$ and $e^-$ beams in
intense laser light. The incoming electrons would then transmit most of
their energy to the photons by inverse Compton scattering. The resulting
photon spectrum is very hard, and the achievable \gaga\ luminosity is similar
to the original \ee\ luminosity. However, we will show that the hadronic
backgrounds at such a \gaga\ collider are much {\em larger} than at \ee\
colliders operating at the same energy. The cross section for the ``direct''
process $\gaga \rightarrow \qqbar$ now falls with energy; however, as mentioned
above, the cross section for resolved photon processes increases with the
incident \gaga\ energy. A harder photon spectrum therefore always implies
larger hadronic backgrounds.

In ref.\cite{10} we used the production of two central jets as benchmark for
hadronic two--photon reactions. Here we present a much more detailed
description of this reaction, including rapidity distributions and
invariant mass spectra. While this is the most common of all hard
two--photon reactions, it is usually not the most important background to
new physics searches, nor the worst obstacle to precision measurements. In
this paper we therefore also give results for a more complete list of hard
two--photon reactions, including events with only one central jet and one
forward jet (mono--jets), heavy quark production, and Drell--Yan
production of $W$ and $Z$ bosons in resolved photon reactions. We find
that, at \ee\ colliders with $\rs \leq 500$ GeV, the production of central
$c \overline{c}$ and $b \overline{b}$ pairs is always dominated by the
direct contribution, except perhaps at very small transverse momenta, $\pT
\leq 5$ GeV. Moreover, the total $t \overline{t}$ cross section at such
colliders will be dominated by the annihilation process. Our calculations
of the annihilation contribution include effects due to initial--state
radiation as well as beamstrahlung; at $\rs = 500$ GeV, these effects
increase (decrease) the annihilation contribution if $m_t < (>)\  155$
GeV.

We also expand our previous discussion \cite{10} of the semi--hard ``minijet''
background. In its simplest form, leading order QCD extrapolated down to
transverse momenta between approximately 1 and 3 GeV predicts an almost
linear increase of the total inelastic \gaga\ cross section with energy.
Of course, this behaviour cannot persist indefinetely. However, we will present
arguments showing that it is at least possible that the mechanism which
ultimately ``unitarizes'' the cross section (e.g., eikonalization) will become
effective only at energies beyond the reach of the next generation of \ee\
colliders. Moreover, we will argue that even an early flattening off of the
cross section need not lead to a sizeable reduction of the total $E_T$ in
the underlying event, which is a good measure for the ``messiness'' of the
environment.

While most of our numerical results will be given for colliders operating
at $\rs = 500$ GeV, which is now envisioned as the likely energy for the
next \ee\ collider \cite{1a}, we also try to extrapolate to higher energies.
We argue that for the so--called mainstream designs utilizing X--band
microwave cavities the occurence of an underlying event, i.e. of multiple
interactions per collision, seems unavoidable at $\rs \geq 1$ TeV,
unless the bunch structure can be modified considerably. On the other hand,
superconducting designs might be able to provide a clean environment up to
$\rs \simeq 2$ TeV.

The rest of this paper is organized as follows. In Sec.2 we present the
necessary formalism. In particular, various parametrizations of the parton
content of the photon are briefly discussed. We also describe the photon
and electron spectra that we use in our calculations. In Sec.3 results for
hard two--photon reactions are presented. We devote different subsections
to the production of di--jets (3a), mono--jets (3b), heavy quarks (3c) and
single $W$ and $Z$ bosons (3d). Sec.4 contains a discussion of the soft and
semi--hard background. Finally, in Sec.5 we summarize our results and present
some conclusions.

\section*{2. Formalism and distribution functions}
\setcounter{footnote}{0}

In this section we describe the techniques necessary to derive the results of
secs. 3 and 4. We will employ the structure function formalism even when
estimating annihilation cross sections. In this formalism, the cross
section for the production of a given final state $X$ is expressed as a
product of the functions $f_1(x_1)$ and $f_2(x_2)$, describing the
probabilities to find particles 1 and 2 with fractional momenta $x_1$ and
$x_2$ in the incident beams, and the hard $1 + 2 \rightarrow X$ scattering
cross section $d \sighat$:
\be \label{e1}
d \sigma = f_1(x_1) f_2(x_2) d \sighat(\shat). \ene
Here $\shat = x_1 x_2 s$ is the invariant mass of the system of particles 1 and
2, and \rs\ is the nominal \ee\ machine energy.

In case of two--photon processes, particles 1 and 2 are either photons, or
quarks or gluons inside a photon. We use the terminology of ref.\cite{6}
to classify the various two--photon processes. In ``direct'' processes,
particle s 1 and 2 are both photons. The only process of this kind which
is of interest to us is the production of a pair of massive or massless
quarks, $\gaga
\rightarrow \qqbar$; the corresponding hard cross section can for instance be
found in ref.\cite{13}. In ``once resolved'' processes (``1--res'' for short)
particle 1 is a photon, while particle 2 is a quark or gluon. The relevant
subprocesses are $\gamma q \rightarrow g q$ and $\gamma g \rightarrow \qqbar$;
their cross sections are listed in ref.\cite{14}. Finally, in ``twice
resolved'' or ``2--res'' processes particles 1 and 2 are both colored
partons. The eight subprocesses contributing to the production of massless
parton jets and their cross sections are given, e.g., in ref.\cite{15}.
The cross sections necessary to compute the production of massive $Q
\overline{Q}$ pairs in once and twice resolved processes can be found in
refs. \cite{16} and \cite{17}, respectively.

We remind the reader at this point that resolved photon processes are
characterized by spectator jets going essentially into the direction of
the incoming photons. These spectator or remnant jets are the result of the
hadronization of the colored system that is produced when a quark or gluon
is taken out of a photon. Although the axes of these jets almost coincide with
the beam directions, at least their outer fringes can emerge at substantial
angles, due to nontrivial color flow between spectator jets and hard jets
as well as the boost from the \gaga\ centre--of--mass system to the lab frame.
For instance, according to the AMY Monte Carlo simulation of their data on
jet production in \gaga\ scattering \cite{7}, about 50\% of the particles
that originate from the spectator jets emerge at angles $\theta > 20^{\circ}$.

We now turn to a discussion of the various probability or distribution
functions $f_i$. In general there are two different contributions to the
photon flux \fge\ in an electron beam. The first contribution is actually an
approximation of the complete two--photon process $\ee \rightarrow \ee X$.
The corresponding cross section can be cast in the form of eq.(\ref{e1})
using the effective photon (EPA) or Weizs\"acker--Williams (WW)
approximation \cite{18}.
We use the expression of ref.\cite{19} to describe the spectrum of
photons that interact directly: \beq \label{e2}
f_{\gamma|e}^{\rm EPA,dir.}(x) = \frac {\aem} {2 \pi x} &\left\{ \left[ 1 +
\left( 1 - x \right)^2 \right] \left( \ln \frac {E^2} {m_e^2} - 1 \right)
\right. \nonumber \\ &\left.
+ x^2 \left[ \ln \frac {2 \left( 1-x \right)}{x} + 1 \right]
+ \left( 2-x \right)^2 \ln \frac {2 \left( 1-x \right)} {2-x} \right\}, \eeq
where $E = \rs/2$ is the nominal electron beam energy and $m_e$ the electron
mass. This expression has been shown \cite{21} to reproduce exact results for
both differential and total cross sections for the two--photon production of
scalars and spin--1/2 fermions to relative accuracy of 10\% or better. However,
it has been derived by integrating the virtuality $- P^2$ of the exchanged
photon over its full kinematically allowed range. On the other hand, it is
known \cite{22} that the parton content of highly off--shell photons is
reduced compared to that of real photons. If the scale $Q^2$ at which the
photon is probed is less than $P^2$, the concept of partons residing ``in''
this photon is no longer applicable; in this kinematical regime the formalism
of deep inelastic scattering should be used, where the charactersistic scale
would be given by $P^2$ rather than $Q^2$. We have conservatively ignored
contributions with $P^2 > Q^2$ altogether. Furthermore we introduce a
numerical suppression factor of 0.85, estimated from results of Rossi
\cite{22}, in order to approximate the suppression of virtual photon
structure functions in the region $\Lambda_{\rm QCD}^2 < Q^2 < P^2$.
Altogether we thus have for the effective spectrum of resolved photons:
\be \label{e3} f_{\gamma|e}^{\rm EPA,res.}(x) = 0.85 \frac {\aem} {2 \pi
x} \left[ 1 + \left( 1 - x \right)^2 \right] \ln \frac {Q^2} {m_e^2}. \ene
We remark that this numerical suppression factor should {\em not} be
introduced if anti--tagging of the scattered electrons already implies
$P^2 \leq \Lambda_{\rm QCD}^2$.

The second major contribution to the photon flux at \ee\ linacs comes from
beamstrahlung \cite{8}. As already mentioned in the Introduction,
beamstrahlung is produced when particles in one bunch undergo rapid
acceleration upon entering the electromagnetic field of the opposite
bunch. The intensity and spectrum of beamstrahlung therefore depend on the
strength and extension of this field, which in turn are determined by the
size and shape of the bunches. Unlike the machine--independent
bremsstrahlung (EPA) contribution described above, the beamstrahlung
contribution to \fge\ therefore depends sensitively on the bunch
parameters of the collider under discussion. In general the relationship
between the photon spectrum and the machine parameters is highly
nontrivial \cite{8,23}. Fortunately, Chen \cite{24} has recently been able
to derive approximate expressions, which accurately reproduce the exact
spectra as long as the fields produced by the bunches are not too strong;
this criterium is always fulfilled for our examples.

In this approximate treatment the beamstrahlung spectrum is determined by
three parameters: The electron beam energy $E$; the bunch length $\sigma_z$
(for a Gaussian longitudinal bunch profile); and the beamstrahlung parameter
\Y, which is proportional to the effective magnetic field of the bunches. For
Gaussian beams, the effective or mean value of \Y\ can be estimated from
\cite{24} \be \label{e4}
\Y = \frac { 5 r_e^2 E N } { 6 \aem \sigma_z \left( \sigma_x + \sigma_y
\right) m_e }. \ene
Here, $N$ is the number of electrons or positrons in a bunch, $\sigma_x$ and
$\sigma_y$ are the transverse bunch dimensions, and $r_e = 2.818 \cdot
10^{-12} \ mm$ is the classical electron radius. Notice that \Y\ decreases
like $\sqrt{ \sigma_y / \sigma_x }$ if $\sigma_x \gg \sigma_y$ with constant
$\sigma_x \cdot \sigma_y$. Moreover, for given luminosity and bunch dimensions,
$N$ is inversely proportional to the square root of the number of bunch
collisions per second; beamstrahlung can therefore be reduced by introducing
more bunches.

In terms of these parameters, the beamstrahlung spectrum can be written as
\cite{24} \beq \label{e5}
f_{\gamma|e}^{{\rm beam}}(x) = \frac {\kappa^{1/3}} {\Gamma(1/3)}
x^{-2/3} \left( 1-x \right)^{-1/3} e^{- \kappa x / \left( 1-x \right)}
&\left\{ \frac {1-w} {\tilde{g}(x)} \left[ 1 - \frac {1}
{\tilde{g}(x) N_{\gamma}}
\left( 1 - e^{-N_{\gamma} \tilde{g}(x)} \right) \right] \right. \nonumber \\
&\left. + \ w \left[ 1 - \frac{1} {N_{\gamma}} \left( 1 - e^{-N_{\gamma}}
\right) \right] \right\}, \eeq
with \be \label{e6}
\tilde{g}(x) = 1 - \frac{1}{2} \left( 1-x \right)^{2/3} \left[ 1 - x +
\left( 1 + x \right) \sqrt{ 1 + \Y^{2/3} } \right] \ene
and $\kappa = 2 / (3 \Y), \ w = 1 / \left( 6 \sqrt{\kappa}\right)$.
Finally, the average number of photons per electron $N_{\gamma}$ is given
by \be \label{e7} N_{\gamma} = \frac {5 \alpha_{\rm em}^2 \sigma_z m_e} {2
r_e E} \frac {\Y} {\sqrt{1 + \Y^{2/3} }}. \ene Eqs.(\ref{e5}) --
(\ref{e7}) are valid as long as \Y\ is not much larger than one,
practically for $\Y \leq 5$ or so.

Notice that the flux of soft photons with $\kappa x \leq 1-x$ actually
decreases slowly with increasing \Y; in contrast, the flux of hard photons
is exponentially suppressed if $\Y \ll x/(1-x)$. Furthermore, we see from
eqs.(\ref{e4}) and (\ref{e7}) that $N_{\gamma}$ is approximately independent
of \rs\ and $\sigma_z$, while $\Y \propto \rs/\sigma_z$; notice that
$f_{\gamma|e}^{{\rm beam}}$ grows almost linearly with $N_{\gamma}$ as long
as $\tilde{g}(x) N_{\gamma} \leq 1$. Increasing the bunch length thus
strongly suppresses the hard part of the beamstrahlung spectrum, but
increases the soft part of the spectrum.

Parameters of some recent designs \cite{1a} of \ee\ linacs are listed in
Table 1. In addition to the nominal centre--of--mass energy, the beamstrahlung
parameter and the bunch length, for future reference
we also give the luminosity per bunch crossing \lhat, the number of bunches
per bunch train $N$, the temporal separation between two consecutive bunches
in a train $\Delta t$, and the total luminosity ${\cal L}$.

The Palmer G and Palmer F designs, first proposed in ref.\cite{25}, as well
as the proposal for the Japan Linear Collider (JLC) \cite{26} all foresee
the use of X--band microwave cavities; these offer strong accelerating fields,
and thus allow to construct relatively short accelerators. In
contrast, the DESY--Darmstadt designs \cite{27,DDP} use larger S--band
cavities; this technology is better understood, but the accelerating
fields are smaller. Finally, the TeV Superconducting Linear Accelerator
(TESLA) design \cite{28} employs superconducting cavities. This allows to
store the microwave energy almost indefinetely, which in turn makes it
possible to use a very large number of well separated bunches with low
luminosity per bunch crossing; we have already seen that this reduces
beamstrahlung.  On the other hand, this design is technologically most
demanding.

The corresponding photon spectra, computed from eqs.(\ref{e5}) -- (\ref{e7}),
are shown in Figs. 1a,b. As expected from the above discussion, the TESLA
beamstrahlung spectrum is very soft; its contribution to the total photon
spectrum is negligible for fractional momentum $x \geq 0.05$. In contrast,
the beamstrahlung spectrum of the Palmer G design is quite hard, dominating
the total photon spectrum out to $x \simeq 0.6$. The other designs fall in
between these two extremes. Notice the cross--over of the beamstrahlung spectra
of the wide band beam (wbb)
DESY--Darmstadt and Palmer F designs; the former uses longer bunches
and thus has more soft beamstrahlung, while the larger \Y\ parameter of the
latter leads to enhanced hard beamstrahlung. Since the hard
contribution to the total photon spectrum is in both cases dominated by the
bremsstrahlung (EPA) contribution, we can expect larger two--photon cross
sections at the DESY--Darmstadt (wbb) collider. The narrow band beam (nbb)
version of this design has a beamstrahlung spectrum which is almost as soft as
that of the TESLA; on the other hand, it has the smallest luminosity of the
designs we studied.

Fig. 1b shows the evolution of the beamstrahlung spectrum at the JLC as the
energy is increased from 0.5 to 1.5 TeV. We see from eq.(\ref{e4}) that,
everything else remaining constant, \Y\ increases linearly with energy.
However, higher energies also necessitate higher luminosities. If this is
achieved by reducing the transverse bunch dimensions $\sigma_x$ and
$\sigma_y \propto 1/\rs$, or by increasing the number of particles per
bunch $N \propto \rs$, \Y\ will grow like $s$, not like \rs. Table 1 shows
that the first planned extension of the JLC, from \rs\ = 0.5 to 1 TeV,
indeed leads to an almost fourfold increase of \Y. The difference between
the corresponding spectra is therefore quite pronounced. In the second step,
from \rs\ = 1 to 1.5 TeV, \Y\ only grows linearly with \rs. The difference in
the spectra, when shown as a function of the scaling variable $x$, is
therefore not very large. This slow increase of \Y, which has been achieved
by increasing the aspect ratio $\sigma_x/\sigma_y$ from about 120 to 200,
implies that the average
number of beamstrahlung photons per electron even decreases \cite{26} in this
second extension, from 1.8 to 1.45; quantum effects, which lead to the
suppression factor $1/\sqrt{ 1 + \Y^{2/3}}$ in eq.(\ref{e7}), also contribute
to the reduction of $N_{\gamma}$.

As already mentioned in the Introduction, it has been suggested \cite{12} to
convert future \ee\ colliders into \gaga\ colliders by backscattering laser
light off the incident electron and positron beams. If these beams are not
polarized, the resulting photon spectrum depends only on the electron energy
and the frequency of the laser. The laser photons should not be too energetic,
since otherwise a backscattered photon and a laser photon could combine to
form an \ee\ pair, which would drastically reduce the electron to photon
conversion efficiency. Here we will assume that the laser frequency is chosen
such that one stays just below this threshold. The spectrum of backscattered
photons is then given by \cite{12} \be \label{e8}
f_{\gamma|e}^{{\rm laser}}(x) = \frac { -0.544 x^3 + 2.17 x^2 -2.63 x +
1.09} { \left( 1-x \right)^2} \cdot \theta(0.828-x). \ene
This spectrum is shown by the long-dashed dotted curve in Fig. 1a; it
slowly rises towards its kinematical cut--off. Notice that the spectrum
(\ref{e8}) expressed in terms of the fractional photon energy $x$ is
{\em independent} of the beam energy; this is a consequence of our assumption
that the energy of the original laser photons decreases like $1/\rs$, as
described above. Since (almost) the total electron beam energy
will be passed on to the photons, there is (almost) no bremsstrahlung
contribution to the photon flux at a \gaga\ collider. We will assume that
at these machines eq.(\ref{e8}) describes the total spectrum.

Of course, in order to compute cross sections for resolved photon processes
one also needs to know the distribution functions $\qvec\xq = (\qig,\Gg)
\xq$ of partons inside the photon, in addition to the photon spectrum. The
$f_i$ of eq.(\ref{e1}) are then given by convolutions of these distribution
functions: \be \label{e9}
f_{\vec{q}|e} \xq = \int_{x}^{1} \frac {dz} {z} \fge(z) \qvec ( \frac {x}{z},
Q^2). \ene
Unfortunately, there is not (yet) much experimental information about
\qvec\xq. The combination $F_2^{\gamma} = x \sum_{i} e_i^2 q_i^{\gamma}$
(up to higher order corrections) has been measured \cite{29} for $x \geq 0.05$
with a precision of typically 10 -- 20 \%; however, these measurements tell us
practically nothing about the flavour structure of the photon, the quark
distributions at low $x$, or the gluon distribution at any $x$. This last
point was demonstrated explicitly in ref.\cite{30}, where it was shown that
very different ans\"atze for $\Gg (x, Q_0^2)$ can lead to almost equally good
descriptions of existing data on $F_2^{\gamma} ( x, Q^2 \geq 4 \ {\rm GeV}^2)$.

We try to give a feeling for the resulting uncertainties by presenting results
for different parametrizations of \qvec\xq. Our ``standard'' choice will
be the ``DG'' parametrization of ref.\cite{31}. It
is free of unphysical $x \rightarrow 0$ divergencies;
it also fits the data on $F_2^{\gamma}$ quite well \cite{30}. The
most important feature of this parametrization relevant for
phenomenological applications is that it assumes that gluons are only created
radiatively in the photon; this leads to a rather soft shape of $\Gg(x)$, as
well as a small total gluon content of the photon.

Our second choice is based on the asymptotic ``DO'' parametrization of
ref.\cite{14}. As discussed in ref.\cite{6}, it has to be augmented by
a ``hadronic'' contribution in order to fit data on $F_2^{\gamma}$ with
$\Lambda_{\rm QCD} = 0.4$ GeV; we estimate \cite{6} this component using Vector
Meson Dominance (VMD) ideas. We use this parametrization mostly to
demonstrate the effects of a relatively hard, truely intrinsic contribution
to \Gg. However, the DO parametrization should not be used at very small
values of $x$, because it suffers from even worse divergencies than the
$x^{-1.6}$ behaviour predicted \cite{3} by the leading order asymptotic
calculation. Since this region is important for the accelerators we
are discussing, we have modified the DO parametrization for small x,
somewhat arbitrarily defined as $x \leq 0.05$: \be \label{e10}
q_i^{\gamma,{\rm mod. DO}} \xq = c_i x^{-1.6} \ln \frac {Q^2}
{\Lambda^2_{\rm QCD}}, \ene
where the $c_i$ are chosen to give smooth transitions at $x = 0.05$; a similar
ansatz has been used for the gluon density. We call the result the ``modified
DO+VMD'' parametrization.

Both the DG and the DO+VMD parametrization are able \cite{7} to describe
the AMY data on jet production quite well, if the minimal partonic
transverse momentum is adjusted properly; we will come back to this point
in sec.4. In contrast, the third parametreization of ref.\cite{30}
(``LAC3'') has been found \cite{31a} to over--estimate the resolved photon
contribution by a large factor. This is due to the extremely hard gluon
density used in this parametrization, $\Gg(x,Q_0^2) \propto x^6$, which
looks quite unnatural.\footnote{The fact that this ansatz reproduces
existing data on $F_2^{\gamma}$ demonstrates once again that these data
give very little information about \Gg.} The other two parametrizations of
ref.\cite{30} are quite similar to each other; we will show some results
for the ``LAC2'' parametrization. However, we again find it necessary to
slightly modify the original parametrization. It gives separate and
different distribution functions for $u, \ d, \ s$ and $c$ quarks; we find
that at small $x$ it usually predicts $c^{\gamma}(x) > u^{\gamma}(x), \
s^{\gamma}(x) > d^{\gamma}(x)$, opposite to the expectation that the
contribution of heavier quarks should be suppressed. We therefore define
effective distribution functions for charge 2/3 and 1/3 quarks: \ben
\label{e11} \beq u_{{\rm eff}}^{\gamma,{\rm LAC}} \xq &= \frac{1}{2}
\left( u^{\gamma,{\rm LAC}} + c^{\gamma,{\rm LAC}} \right) \xq;
\label{e11a} \\ d_{{\rm eff}}^{\gamma,{\rm LAC}} \xq &= \frac{1}{2} \left(
d^{\gamma,{\rm LAC}} + s^{\gamma,{\rm LAC}} \right) \xq. \label{e11b}
\eeq \een
Note that the DG and DO+VMD parametrizations also assume $s^{\gamma} =
d^{\gamma}$ and $c^{\gamma} = u^{\gamma}$ (for $N_f \geq 4$).

Very recently, two more sets of parametrizations of \qvec\ have been proposed.
In ref.\cite{GRV} Gl\"uck et al. give a parametrization of their ``dynamical''
prediction for the photon structure function \cite{dyn}. They assume a hard,
valence--like gluon distribution at a very low input scale $Q_0 = 300$ MeV.
As a result their \Gg\ resembles the DO+VMD parametrization at median and
large Bjorken--$x$ and low $Q^2$, but becomes more similar to the DG
parametrization for low $x$ and/or high $Q^2$. In contrast, Gordon and
Storrow \cite{GS} use a rather high input scale;
their input is a sum of a VMD part and a ``pointlike" part estimated from the
quark--parton model. They give two parametrizations, depending on whether
gluon radiation from the ``pointlike" part of the quark densities is included.
At median and large $x$ and low $Q^2$ their gluon densities lie between those
of the DG and DO+VMD parametrizations. At low $x$ and low $Q^2$ it falls
even below the DG prediction for \Gg; however, this is mostly due to their
choice of a rather high value of $Q_0$. In fact, their parametrization
cannot be used for $Q^2 < 5.3 \ {\rm GeV}^2$, so that it cannot predict total
\ccbar\  (sec. 3c) or minijet (sec. 4) cross sections. In any case, by
comparing predictions from the DG, modified DO+VMD and LAC2 parametrizations
we still span the whole range of existing parametrizations for \qvec, with
the parametrizations of refs.\cite{GRV,GS} falling somewhere in between.

We will give results for processes characterized by momentum scales between
a few and a few hundred GeV. In between, two flavor thresholds are crossed.
The problem of heavy quark distribution functions in the photon still awaits
a rigorous treatment \cite{31b}. For simplicity we will assume $N_f = 3$
massless flavors in the photon if the momentum scale $Q^2 < 50 \ {\rm GeV}^2, \
N_f = 4$ for $50 \ {\rm GeV}^2 \leq Q^2 \leq 500 \ {\rm GeV}^2$,
and $N_f = 5$ for
$Q^2 > 500 \ {\rm GeV}^2$. We use the interpolating expression of ref.\cite{32}
for $\alpha_s$, with $m_c = 1.5$ GeV, $m_b = 5$ GeV and $m_t = 100$
GeV.\footnote{The variation of $\alpha_s$ with $m_t$ is always negligible
for our processes.} When using the DG or modified DO+VMD parametrizations
we assume $\Lambda_{\rm QCD} = 0.4$ GeV, while the LAC parametrizations have
to be used with $\Lambda_{\rm QCD} = 0.2$ GeV.

As mentioned at the beginning of this section, we will employ the structure
function formalism of eq.(\ref{e1}) also to compute annihilation cross
sections.  In this case we will use it to include the effects of initial
state radiation and beamstrahlung. Both effects smear out the electron
distribution function $\fee (x)$ from the ideal $\delta$--function at
$x=1$. Initial state radiation (ISR) is described by (to one loop order)
\cite{33} \be f_{e|e}^{{\rm ISR}}(x) = \frac {\beta} {2} \left( 1-x
\right)^{\beta/2 - 1}
\left( 1 + \frac{3}{8} \beta \right) - \frac{1}{4} \beta \left( 1 + x
\right), \label{e12} \ene
where \begin{eqnarray}
\beta = \frac {2 \aem} {\pi} \left( \ln \frac {s} {m_e^2} - 1 \right) .
\nonumber \end{eqnarray}
The first term in eq.(\ref{e12}) re--sums leading logarithms near $x=1$, i.e.
includes soft photon exponentiation. Numerically, $\beta = 0.124$ at \rs =
500 GeV. Even at this high energy the electron spectrum (\ref{e12}) is
strongly peaked at $x=1$, with 79\% (26\%) of all electrons satisfying
$x > 0.99 \ (0.999)$. Notice that eq.(\ref{e12}) satisfies the charge
conservation constraint $\int_0^1 \fee (x) = 1$ exactly.

We again use the approximate formalism of ref.\cite{24} in order to describe
the effects of beamstrahlung. Here the electron spectrum is given by the
function $\psi$: \beq \label{e13}
\psi(x) = \frac {1} {N_{\gamma}} & \left\{ \left( 1 - e^{- N_{\gamma}}
\right) \delta (1-x)
\right. \nonumber \\ &\left.
+ \ \frac {e^{- \eta(x)}} {1-x} \sum_{n=1}^{\infty} \left[ \left( 1-x
\right) + x \sqrt{ 1 + \Y^{2/3} } \right]^n \frac {\eta(x)^{n/3}}
{n! \Gamma(n/3)} \gamma(n+1,N_{\gamma}) \right\}, \eeq
where $\eta(x) = \kappa \left( 1/x - 1 \right)$, and $\gamma(a,b)$ is the
incomplete $\gamma$--function, for which we use a power expansion \cite{34}.
The parameters $N_{\gamma}$ and $\kappa$ have already been introduced in the
discussion of the beamstrahlung photon spectrum. Notice that $\psi(x)$ is
approximately, but not exactly normalized to 1. The fact that the deviation
is always less than 10\% for the machines we are considering gives us some
confidence that the formalism of ref.\cite{24} is indeed applicable to them.
Nevertheless one would like to achieve better than 20\% precision at least
for annihilation cross sections. We thus write \be \label{e14}
f_{e|e}^{{\rm beam}}(x) = \frac {\psi(x)} {\int_0^1 \psi(z) dz}, \ene
which satisfies charge conservation exactly. Another possibility would have
been to adjust the coefficient of the $\delta$--function in eq.(\ref{e13})
such that charge is conserved exactly. Since most of the electron spectrum
is concentrated at and just below $x=1$, the difference between these two
procedures is very small.

Beamstrahlung and initial state radiation are characterized by quite different
time or length scales. The final electron spectrum can therefore to very
good approximation be obtained by simply convoluting eqs.(\ref{e12})
and (\ref{e14}): \be \label{e15}
\fee (x) = \int_x^1 \frac {dz} {z} f_{e|e}^{{\rm ISR}}(z)
f_{e|e}^{{\rm beam}}( \frac {x}{z}). \ene
Since the calculation of $\psi(x)$, eq.(\ref{e13}), is numerically quite
costly, we found it convenient to use a cubic spline interpolation for
$\fee(x)$.

The resulting electron spectra for some designs of \ee\ colliders
operating at \rs\ = 500 GeV are shown in fig. 2. For comparison we also
show a curve where beamstrahlung is not included, so that the spectrum is
simply given by eq.(\ref{e12}) (dotted line). We see that at the TESLA
design, beamstrahlung affects the spectrum only in the region $x \geq 0.95$;
while this may still have some impact on the study of new thresholds, it is
unimportant for our purposes, since our cross sections do not depend very
sensitively on the incident electron energy.
In contrast, at the Palmer G design beamstrahlung
modifies the electron spectrum at {\em all} values of $x$; it increases
$\fee(x=0.5)$ by more than an order of magnitude. For all the other
designs for colliders operating at this energy the low energy end of the
electron spectrum is essentially given by the bremsstrahlung contribution
alone. This is true even for the JLC1 design, which has the second largest
beamstrahlung contribution of the designs we studied. Notice that
for machines with little or no beamstrahlung, a substantial part of
the integral over \fee\ comes from the region $x >$ 0.99, which is not shown
in fig. 2.

Finally, we mention that we used a running electromagnetic coupling constant
when computing annihilation cross sections: \be \label{e16}
\alpha_{{\rm em}}^{-1} = 128 \left( 1 - \frac {20} {9 \pi}
\frac {1} {128} \ln \frac {\hat{s}} {m_Z^2} \right). \ene
This expression includes contributions from all light fermions, including
$b$--quarks, but no $t$ or $W$ loops. Numerically, \aem (1 TeV) = 1/124.6.
Of course, we include both $\gamma$ and $Z$ exchange contributions to the
annihilation cross section. We do, however, not include QCD corrections,
since this would be quite nontrivial in case of the two--photon cross
sections.\footnote{Of course, in many cases the QCD corrections to annihilation
cross sections can be estimated by simply multplying the cross section with
$1 + \alpha_s/\pi$, leading to a 3 to 5\% increase of the cross section.} The
distinction between direct and resolved contributions
becomes blurred in higher orders; e.g., QCD corrections to the direct
process contain collinear divergencies which have to be absorbed in the
parton distribution functions \cite{fritz}.
Our results for annihilation cross sections should therefore be precise to
about 5 to 10\%.

We are now in a position to present our numerical results. We start with a
discussion of various hard two--photon induced backgrounds.

\section*{3. Hard two--photon reactions}
\setcounter{footnote}{0}
In this section results for ``hard'' two--photon processes are given, the
cross sections of which can in principle be calculated unambiguously from
perturbative QCD once the parton densities inside the photon are known.
By far the most common of these processes is the production of two
high--\pT\ jets. If both jets are produced at large angles, this process
leads to a di--jet final states, which we discuss in sec. 3a; in sec. 3b we
present results for the case that one of the two jets emerges at a very
small angle, which leads to mono--jet events. The production of heavy quarks
($c \overline{c}, \ b \overline{b}$ and $t \overline{t}$) is discussed in
sec. 3c. In sec. 3d the production of single $W$ and $Z$ bosons is studied;
these events are comparatively rare, but offer quite striking signatures if
the gauge bosons decay leptonically.

\subsection*{3a. Di--jet production}
This reaction offers the largest rates of all hard, hadronic two--photon
processes; it is also the only one for which experimental data have been
analyzed \cite{35,7}. As already described in sec. 2, all three classes of
two--photon production mechanisms (direct, once resolved and twice resolved)
contribute here. Recall that the once (twice) resolved contributions are
characterized by one (two) spectator jets in addition to the high--\pT\
jets. However, since the axes of these jets coincide essentially with the
beam pipes, it will most likely not be possible to measure their energy on
an event-by-event basis.\footnote{This does not contradict our previous claim
that perhaps as many as 50\% of all particles originating from those jets
will emerge at large angles. The average transverse momentum of particles
from the spectator jets will be a few hundred MeV, while their longitudinal
momentum can be many GeV; the most energetic particles will therefore emerge
at the smallest angle, and thus escape detection.} The only inclusive
observables are thus the transverse momenta and angles or rapidities of the
two high--\pT\ jets. In the leading logarithmic approximation of eq.(\ref{e1}),
the transverse momenta of both jets are equal and opposite; this is exactly
true for on--shell (beamstrahlung) photons, and should still hold to good
approximation for bremsstrahlung photons, due to the $1/Q^2$ behaviour of the
photon propagator. Any given event can thus be characterized by the three
variables $\pT, \ y_1$ and $y_2$, where $y_i$ denotes the rapidity of the
$i$--th jet. On the parton level, these three observables are related to
the fractional momenta $x_i$ of eq.(\ref{e1}) via \ben \label{e17} \beq
x_1 &= \frac {x_T} {2} \left( e^{y_1} + e^{y_2} \right) ; \label{e17a} \\
x_2 &= \frac {x_T} {2} \left( e^{-y_1} + e^{-y_2} \right), \label{e17b}
\eeq \een
where \be \label{e18}
x_T = \frac {2 \sqrt{ p_T^2 + m^2} } { \sqrt{s}} \ene
is an ``average'' or ``typical'' value for the $x_i$. The Mandelstam variables
$\hat{t}$ and $\hat{u}$ of the $2 \rightarrow 2$ subprocess are given by
\be \label{e19}
\hat{t}, \ \hat{u} = m^2 + \frac { \hat{s} } {2} \left( -1 \pm \sqrt
{1 - x_T^2 } \right). \ene
For future reference we have allowed for a finite (equal) mass $m$ of the
two produced partons; in this and the next subsection we will be concerned
with the case $m=0$.

In this subsection we require both jets to be produced
centrally. In this context it is important to realize that detectors at
future \ee\ linacs will almost certainly have substantial dead areas around
the beam pipes, i.e. will not be very hermetic. The reason is that
beamstrahlung also gives rise to enormous numbers of \ee\ pairs \cite{36,23}.
For instance, at the Palmer G design, one might have to expect \cite{9} about
500,000 such pairs per bunch crossing. Fortunately, a large majority of these
electrons will be produced at small angles and with small transverse momentum;
the central part of the detector should therefore remain relatively free of
these electrons, although a few central pairs per bunch crossing might still
have to be expected \cite{23,37}. However, the large electron flux at small
angles will make it almost impossible to extend the detector close to the
beam pipe. We will assume that the angular coverage for jets only extends
out to $\theta = 15^{\circ}$, which corresponds to \be \label{e20}
|y_{1,2}| \leq 2. \ene

In order to give a first idea of the magnitude of hard two--photon cross
sections, we show in fig. 3a the total cross section for the production of
two central jets with $\pT \geq \ptmin$, as a function of \ptmin, for the
first four machines of table 1 as well as the \gaga\ collider.
For fig. 3 and the remaining figures of this subsection we have used
$Q^2 = \hat{s}/4$ as the scale in the parton densities,
including the bremsstrahlung spectrum of resolved photons (eq.(\ref{e3})),
as well as in $\alpha_s$; choosing $Q^2 = p_T^2$ instead would have changed
the results only by about 10\%, but would have lead to even more pronounced
kinks at $\pT = \sqrt{50}$ and $\sqrt{500}$ GeV, where the number of
participating flavors is changed, as described in sec. 2. The results of fig. 3
have been obtained using the DG parametrization;
as will be discussed in more detail later,
the other parametrizations mentioned in sec. 2 would lead to even larger
cross sections. Notice that the QED point cross section $4 \pi
\alpha^2_{\rm em} / (3 s)$ only amounts to 0.4 pb at \rs\ = 500 GeV. The
two--photon cross section even for quite hard jets ($\pT > 10$ GeV) is
between 50 and 1,000 times larger, where the smaller (larger) number refers
to the TESLA (Palmer G) design. The luminosity per year of $10^7$ seconds
varies between 14 $fb^{-1}$ at Palmer F and 60 $fb^{-1}$ at Palmer G.
The two--photon contribution to the total di--jet rate should therefore
in principle be measurable out to $\pT = 150$ GeV at least. More importantly,
one expects between 4 (at TESLA) and 250 (at Palmer G) {\em million}
two--photon events per year with $\pT > 5$ GeV.

We see from fig. 1 that at very large \pT\, all \ee\ colliders must have the
same two--photon cross sections, since $\fge(x)$ is dominated by the
bremsstrahlung (EPA) contribution as $x \rightarrow 1$. Indeed, at the TESLA
and DESY-Darmstadt (nbb)
colliders beamstrahlung increases the total di--jet production with $\pT > 20$
GeV only by 20\% or less; for the DESY--Darmstadt (wbb),
Palmer F and JLC machines
this is true only for $\pT > 75$ to 100 GeV. Finally, at the Palmer G design,
the beamstrahlung contribution remains sizeable at all values of \pT\ where
the di--jet cross section is measureable. For all \ee\ colliders the cross
section falls quite rapidly with \ptmin. In contrast, the very hard photon
spectrum of the \gaga\ collider leads to a relatively flat \pT\ spectrum
once $\pT > 50$ GeV or so; here the total rate is dominated by the direct
process $\gaga \rightarrow \qqbar$.

In fig. 3b we show the integrated di--jet cross section for the three stages
of the JLC. We also give a first indication of the relative importance of
the various contributing processes by showing separate curves for the direct
process (dashed) and the total cross section (solid). The evolution of the
direct cross section with energy closely follows that of the photon spectrum,
see fig. 1b: At small \pT, corresponding to small $x$, the cross section
decreases with energy, due to the depletion of soft photons when \Y\ is
increased; the cross section at high \pT\ increases quite rapidly with
energy, since lager \Y\ lead to a rapid increase of the flux of hard photons.

However, fig. 3b also shows that at small \pT, the total cross section is
dominated by resolved photon contributions. Recall that their cross sections
increase \cite{6} with the \gaga\ centre--of--mass energy \wgg. Therefore
events with large \wgg\ make sizeable contributions even in the region of
rather small \pT, in spite of the decrease of the photon flux with \wgg. As
a result, the total di--jet rate increases monotonously with energy for all
\pT. Moreover, the region where resolved photon processes dominate increases
with increasing energy, even when this region is expressed in terms of the
scaling variable $x_T$ of eq.(\ref{e18}).

This discussion shows that harder photon spectra favour resolved photon
processes compared to the direct process. For instance, at the TESLA collider
with its very soft beamstrahlung photon spectrum, resolved photon contributions
dominate \cite{37} total di--jet production only for $\pT \leq 5$ GeV, while
at the \gaga\ collider they remain dominant up to $\pT \simeq 50$ GeV. We also
find that the once resolved contribution exceeds the twice resolved one only
for those values of transverse momentum where the
total rate is already dominated by the direct
process. This is because the twice resolved contribution gets a dynamical
enhancement factor \cite{6} $\hat{s}/\hat{t} \propto \hat{s}/p_T^2$ compared
to both the direct and the once resolved contributions; the former can
proceed via gluon exchange in the $t$ channel, leading to a $1/\hat{t}^2$
pole in the matrix elements \cite{15}, while the latter only have $1/\hat{t}$
poles in the matrix elements, originating from $t$ channel quark exchange.
However, we will see below that the once resolved contribution {\em can} be
dominant in certain kinematical configurations.

In fig. 4 we give more details about the final state composition of the once
(fig. 4a) and twice (fig. 4b) resolved contribution; in these and the
following figures, $q$ stands for a quark or antiquark of any flavour. Not
surprisingly, we see that final states that require a gluon in the initial
state ($qq$ in the 1--res contribution, and $qg$ and $gg$\footnote{We also
include the contribution from $\qqbar \rightarrow gg$, but it is always
very small.} for the 2--res contribution) have a steeper \pT\ spectrum than
those that originate from purely quarkonic initial states; we have already
seen in sec. 2 that all reasonable parametrizations of the parton densities
inside the photon predict $\Gg(x)$ to be much softer than the $\qig(x)$.
Notice that the $qg$ final state makes an important contribution over a
wide range pf \pT\ values. The hard quark distribution functions allow to
probe the gluon density at quite small $x$, where it is large. Moreover,
the hard $qg \rightarrow qg$ matrix element is dynamically enhanced \cite{15}
by a color factor of 9/4 compared to $qq \rightarrow qq$ matrix elements.

Note that the relative importance of the various final states depends on the
photon spectrum in a nontrivial way. We have already seen that harder photon
spectra generally favour more resolved processes. Since harder photons
allow to probe the parton densities inside the photon at small Bjorken--$x$,
see eq.(\ref{e9}),
they also favour gluon--initiated processes over quark--initiated ones. One
must realize, however, that the addition of even a relatively hard
beamstrahlung spectrum, like the one at the JLC for which the results of
fig. 4 have been obtained, can lead to an effectively softer shape for the
total photon spectrum. This is because bremsstrahlung always dominates in
the limit $x \rightarrow 1$; beamstrahlung can only add to the (comparatively)
soft part of the photon spectrum. Moreover, as discussed above, the three
classes of processes (direct, 1--res and 2--res) get contributions from quite
different parts of the photon spectrum, for a given value of \pT. For instance,
at the TESLA collider gluon--induced processes never dominate the total single
resolved contribution, which are dominated by events with quite small \wgg,
where the TESLA photon spectrum is soft, due to the very soft beamstrahlung
spectrum. However, the twice resolved processes, especially those involving
gluons in the initial state, are dominated by events with much larger
\wgg, where the beamstrahlung contribution is already negligible at TESLA;
the total photon spectrum in this region is dominated by the hard
bremsstrahlung contribution of eq.(\ref{e3}). As a result, the cross--over
between the $qq$ and $qg$ final states within the 2--res contributions
occurs at {\em larger} \pT\ at TESLA (28 GeV) than at the first stage of the
JLC (18 GeV) or even Palmer G (20 GeV). Of course, the cross sections for all
final states increase quite rapidly when going from TESLA over JLC1 to
Palmer G, as shown in fig. 1; however, the above discussion shows that
the cross sections for the various subclasses of contributions increase
at quite different rates when the beamstrahlung spectrum is made harder.
Finally, at the \gaga\ collider with its very hard photon spectrum, the
cross--over between quark--initiated and gluon--initiated processes only
occurs \cite{37} at $\pT \simeq 45$ GeV.

Of course, the relative importance of the various initial and final state
configurations also depends on the parton distribution functions \qvec\xq.
We mentioned already in sec. 2 that the DO+VMD parametrization predicts quite
similar quark distribution functions inside the photon as the DG
parametrization does\footnote{Except for the region $x < .1$; however,
here gluon initiated processes overwhelm quark initiated ones anyway.},
while its
\Gg\ exceeds that of the DG parametrization by roughly a factor of 2. It
thus predicts
\cite{37} approximately two times larger rates for the 1--res $qq$ and 2--res
$qg$ final states, and a four times larger rate for the $gg$ final state.

The differences predicted by the recent parametrization of ref.\cite{30} are
even larger, as shown in fig. 5a,b; here we show results for the least
extreme\footnote{The LAC1 parametrization uses an even steeper gluon
distribution function. We have already seen in sec. 2 that the LAC3
parametrization, whose gluon density peaks at large $x$, is strongly
disfavoured by the AMY data \cite{7} on jet production.}
of the three LAC parametrizations, normalized to the prediction of the
DG parametrization. At very small $x$ and small $Q^2$, its gluon
density is about 7 times larger than that of the DG parametrization.
This far over--compensates the reduction of $\alpha_s$ which is induced
by the reduction of $\Lambda_{\rm QCD}$
from 0.4 GeV (DG) to 0.2 GeV (LAC); this reduction
amounts to a factor of 0.7 (0.5) for the 1--res (2--res) processes at \pT\ = 2
GeV. This
manifests itself in the 1--res $qq$ final state, which comes from a $g \gamma$
initial state; the hard photon spectrum leads to a very small average x for
the gluon inside the other photon. If both initial state particles are
gluons, their average Bjorken $x$ has to be increased; therefore the
enhancement factor for the $gg$ final state at the smallest transverse
momentum shown is not 25, but ``only'' 15. The enhancement factors for both
the 1--res $qq$ and the $gg$ final states eventually flatten out when one
goes to even smaller $x$, which can be achieved by using a harder photon
spectrum; e.g., at the \gaga\ collider they reach 4.8 and 25, respectively,
at \pT\ = 2 GeV.

In sharp contrast, the effective
quark density at small $x$ and small $Q^2$ predicted by the LAC2
parametrization appears to be only 70\% of that of the DG parametrization.
This is because the authors of ref.\cite{30} treat the charm quark as a
massless parton already at $Q^2 = 4 \ GeV^2$; however, we do not include
the contribution from $c$ quarks if $Q^2 < 50 \ GeV^2$. Without the $c$ quark
contribution, the LAC parametrization cannot reproduce data on $F_2^{\gamma}$
at small $Q^2$; one might therefore argue that for $Q^2 < 50 \ GeV^2$, we
should have defined the effective LAC $u$ quark density as the sum of the
original $u$ and $c$ quark densities, rather than as the arithmetic mean as
shown in eq.(\ref{e11a}).\footnote{However, in this case it is not clear
how the sizeable contribution from $\gamma \gamma^* \rightarrow c \bar c$ to
$F_2^{\gamma}$ should have been treated.} In that case the predictions of
the LAC2 parametrization for purely quark initiated processes would have been
quite similar to that of the DG parametrization. Notice that the 2--res $qq$
final state in fig. 5b shows no depletion at small \pT; the reason is that the
LAC2 parametrization predicts a sizeable contribution to this final state
from $gg$ fusion, inspite of the smallness of the hard $gg \rightarrow \qqbar$
matrix element compared to the one for $qq \rightarrow qq$ \cite{15}. In any
case, we have already seen that even the DG parametrization with its small
gluon density predicts quark--initiated processes to be sub--dominant for
$\pT < 5$ GeV, so that this discussion is somewhat academic.

In view of these very large differences in the region of small \pT, and
correspondingly small $x$ and small $Q^2$, it is reassuring to note that the
two parametrizations make quite similar predictions both for quark initiated
and for gluon initiated processes once $\pT > 20$ GeV, which corresponds to
$Q^2 > 400 \ GeV^2$ and average Bjorken $x$ for the parton in the photon
larger than 0.15. Due to the increase of $\qvec\xq \propto \log Q^2$, the
ansatz one assumes for \qvec\ at $Q^2 = Q_0^2 = 1$ to $ 4 \ GeV^2$ has only
little effect in this kinematical region, although deviations by 20--30\%
are still possible, e.g. due to the different values for $\Lambda_{\rm QCD}$
that have been assumed. This result also holds for the \gaga\ collider, as far
as the twice resolved contributions are concerned. However, due to the
hardness of the photon spectrum and resulting small average Bjorken $x$ in
the 1--res $qq$ final state, the prediction from the
LAC2 parametrization still exceeds the one
from DG by a factor of 2 at \pT\ = 20 GeV; the two parametrizations make
approximately equal predictions only for $\pT > 40$ GeV. Finally, we remark
that the use of the DO+VMD parametrization at such large \pT\ and
correspondingly large $Q^2$ can be dangerous, since it assumes a $Q^2$
independent (scaling) VMD contribution, in contradiction to expectations
from QCD that the assumed hard gluon component should ``shrink'' down to
small values of $x$.

More detailed information about two--photon contributions to di--jet production
can be gained from the triple differential cross section $d \sigma / d p_T
d y_1 d y_2$. In fig. 6 we display predictions for this quantity as derived
from the DG parametrization, at fixed \pT\ = 30 GeV for the case $y_1 = y_2
 \equiv y$; fig. 6a is for the TESLA collider,
while fig. 6b shows results for the
\gaga\ collider. Only the region $y \geq 0$ is shown; the distributions are
symmetric in $y$, of course.

The shape of the curves can be understood from the observation that increasing
$y$ increases the Bjorken--$x$ of one parton inside the electron, $x_1$,
while decreasing the other, $x_2$; see eqs.(\ref{e17}). The requirement
$x_1 \leq x_{\rm max}$ immediately gives \be \label{e21}
y \leq y_{\rm max} \equiv \log \frac {x_{\rm max}} {x_T}. \ene
For an \ee\ collider, $x_{\rm max} = 1$, while for the \gaga\ collider,
$x_{\rm max} = 0.828$, see eq.(\ref{e8}); therefore the curves in fig. 6b end
at a somewhat smaller value of $y$ than those in fig. 6a. In the limit $y
\rightarrow y_{\rm max}$, we have $x_1 \rightarrow x_{\rm max}$ and
$x_2 \rightarrow x_T^2 / x_{\rm max} \simeq 0.014$ (0.017) at \ee\ (\gaga)
colliders, for the given values of \pT\ and \rs. The region of large $y$ is
therefore sensitive to both the photon density ``in'' the electron and the
parton densities inside the photon at quite small Bjorken--$x$, even at this
large value of \pT\ (which correpsonds to annihilation events at the
TRISTAN collider).

We saw already in fig. 1a that, due to the beamstrahlung contribution,
the TESLA photon spectrum increases rapidly in
the region of small $x$; fig. 6a shows that this leads to an increase of
the direct contribution at large $y$. This shows that one cannot
ignore the beamstrahlung contribution even though it increases the di--jet
cross section integrated over rapidities $|y_{1,2}| \leq 2$ by only
approximately 15\%; without this contribution, the rapdity distribution
would have the bell shape familiar \cite{6} from lower energy colliders.
Of course, at most one of the two initial state photons at large $y$ will come
from beamstrahlung, since $x_1$ is large here; notice that the bremsstrahlung
spectrum (\ref{e2}) remains finite as $x \rightarrow 1$. The direct
contribution at the \gaga\ collider also increases as $y$ approaches its
kinematical maximum. In this case, however, this is due to the increase of
\fge\ at large $x$; fig. 1a shows that it remains essentially constant as
$x \rightarrow 0$.

The once resolved contribution also remains finite as $y \rightarrow
y_{\rm max}$. It is important to realize that in this case only the product
of the photon energy and the Bjorken--$x$ of the parton inside the photon is
fixed, as shown by eq.(\ref{e9}). The once resolved contribution at TESLA
remains large at large $y$ mostly due to the contribution of hard quarks in
soft photons. In contrast, the enormous spike\footnote{This is an example
where the once resolved contribution dominates, at least in a limited region
of phase space.} at large $y$ predicted for the \gaga\ collider is entirely
dominated by soft gluons and sea--quarks in hard photons. This difference
also manifests itself in the energy of the spectator jet, which for
$y \rightarrow y_{\rm max}$ always emerges at negative rapidities, well
separated from the high--\pT\ jets. At $y=2$, the DG parametrization predicts
the average energy of this jet at TESLA
to be 57 GeV, while at $y=1.8$ at the \gaga\
collider it should be as large as 135 GeV. The difference in
spectator jet energy between the two
1--res final states at the TESLA collider is even larger: The $qg$ final
state only has an average spectator jet energy of 31 GeV, while the $qq$
final state, which originates from a (soft) gluon in the initial state,
is accompagnied by a spectator jet with average energy around 100 GeV.
This large difference should be observable even in a detector with
relatively poor angular coverage.

Finally, the rapidity distribution of the twice resolved contribution always
has a maximum at $y=0$. Since $\qvec (x \rightarrow 1) \rightarrow 0$, this
contribution always vanishes as $y$ approaches its kinematical maximum. In
principle, this does not exclude the possibility of having a maximum at
intermediate values of $y$. Indeed, such a maximum does occur at the \gaga\
collider for the 2--res $qg$ final state; here the asymmetric initial state
favours configurations where a hard quark scatters off a soft gluon. This
maximum, which occurs at $y \simeq 1.4$, explains why the total twice resolved
contribution shows a very flat rapidity distribution almost all the way out
to the kinematical maximum.

We close this subsection with a comparison of two--photon and annihilation
contributions to di--jet production at \ee\ colliders with \rs\ = 500 GeV. In
fig. 7 we present the di--jet invariant mass distributions for the two most
extreme examples of table 1, the Palmer G (fig. 7a) and TESLA (fig. 7b)
designs. The contributions from the three subclasses of two--photon
contributions are shown separately, and compared to the annihilation
contribution (dotted curves); both beamstrahlung and initial state radiation
have been included for the latter, using eqs. (\ref{e13}) -- (\ref{e15}).

The most prominent feature of the annihilation contribution is the peak at
$M_{jj} = m_Z$. The annihilation spectrum is quite flat between about 130
and 300 GeV, since the reduction from the $s$--channel propagators is largely
cancelled by the increase of the \ee\ flux with increasing invariant mass.
Of course, at both machines one finds a second, pronounced maximum at large
$M_{jj}$, close to the nominal \rs\ of the machine. The shoulder in the TESLA
annihilation contribution at $M_{jj} \simeq 70$ GeV occurs since requiring
rapidities $|y_{1,2}| \leq 2$ and $M_{jj} < \rs \cdot e^{-2}$ is inconsistent
with $x_1 = 1$ or $x_2 = 1$, see eqs.(\ref{e17}); such events can thus only
occur if {\em both} the electron and the positron emit a hard photon before
annihilating each other. Fig. 2 shows that the
\ee\ flux at TESLA is little affected by
beamstrahlung in the region $M_{jj} < 400$ GeV or so; hard bremsstrahlung
only occurs with probability $\aem/\pi \log E/m_e \simeq 0.03$, so that
double bremsstrahlung is much less likely than single bremsstrahlung. This
shoulder is not visible for the Palmer G design, since here beamstrahlung
affects the \ee\ flux at all invariant masses.

It is obvious, however, that beamstrahlung affects the two--photon contribution
{\em much} more than the annihilation
contribution. While the rate of two--photon
events shoots up by about a factor of 35 when going from TESLA to Palmer G,
the annihilation cross section at $M_{jj} \simeq m_Z$ only increases by a
factor of 3. In fig. 7a we have chosen
the cut $\pT > 20$ GeV, which reduces the two--photon contribution at $M_{jj}
\simeq m_Z$ by
about 25\%, while leaving the $Z$ signal essentially unaltered. An optimal
signal--to--background
ratio can probably be achieved by choosing a cut around 30
GeV; applying an even stronger cut might not help much, since one
then starts to loose significant numbers of annihilation events, partly due
to mismeasurement of the true \pT. With the requirement $\pT > 30$ GeV, the
di--jet annihilation cross section integrated over 87 GeV $\leq \ M_{jj}
\leq$ 95 GeV at Palmer G becomes 6.0 pb, compared to a two--photon background
of 2.9 pb. With the same cuts, the annihilation and two--photon cross sections
at the first stage of the JLC are 2.6 pb and 0.38 pb; the corresponding
numbers for TESLA are 2.0 and 0.09 pb. Predictions for the DESY--Darmstadt and
Palmer F designs fall in between those for TESLA and JLC1.

In fig. 7b we have chosen a very loose explicit \pT\ cut; note, however, that
the rapidity cuts imply $\pT > M_{jj}/8$. Nevertheless this figure nicely
demonstrates the effect of the dynamical enhancement factor $\hat{s}/\hat{t}
\propto M^2_{jj}/p_T^2$ of the twice resolved contribution, which we already
discussed in connection with fig. 3b. We have seen that in the \pT\ spectrum
of di--jet events at TESLA, resolved photon contributions dominate only for
$\pT < 5$ GeV, and that the cross--over between the once and twice resolved
contributions occurs at \pT\ = 28 GeV. From naive kinematical considerations
one would therefore expect the resolved photon contributions to dominate only
for $M_{jj} \ < $ 10 GeV, while fig. 7b shows that they actually are
dominant up to $M_{jj}$ = 60 GeV; similarly, the cross--over between 1--res and
2--res contributions occurs at $M_{jj} \simeq 200$ GeV, which is three times
the value one would expect from kinematics alone, given the \pT\ spectrum.
This enhancement factor also implies that twice resolved contributions will
be even more strongly suppressed by a tight cut on \pT\ than the other
two--photon contributions; this can be seen from fig. 7a, where the twice
resolved contribution is always below the once resolved one.

This figure also shows that in the region $M_{jj} \geq m_Z$, the total
two--photon background is dominated by the direct contribution, once a modest
cut on \pT\ has been applied; it is therefore almost independent of the
parton densities \qvec. We can thus conclude with some confidence that at a
machine like TESLA or the narrow band beam version of the DESY--Darmstadt
design one can study the process $\ee \rightarrow \qqbar$ down to an
invariant mass of about 85 GeV, with little backgound from two--photon
reactions. This might offer the possibility to directly measure \cite{38}
the running of $\alpha_s$ in a single experiment, by comparing
annihilation events at $M_{jj} \simeq m_Z$ with those at $M_{jj} \simeq
\rs$. At the intermediate machines (DESY--Darmstadt (wbb), Palmer F and
JLC1) this should still be possible, but at Palmer G a substantial
irreducible two--photon background will remain.

\setcounter{footnote}{0}
\subsection*{3b. Mono--jet production}
So far we have only considered the case where both high--\pT\ jets are produced
centrally. In this section we discuss the case where only one jet is produced
centrally, while the other is produced at small angles and thus cannot be
reconstructed. To be specific, we require \ben \label{e22} \beq
|y_1| &\leq 1.5; \label{e22a} \\
|y_2| &\geq 2, \label{e22b} \eeq \een
i.e. we demand a finite rapidity gap between the two jets. Since most of the
forward jet will not be seen,\footnote{Part of this jet should still be
visible in most cases; the arguments for the detectability of the spectator
jets in resolved two--photon events also apply here.} the \pT\ of the central
jet will be approximately equal in magnitude to the total missing \pT\ in
the event. Missing \pT\ is (part of) the signature for many interesting
annihilation events. Within the standard model, these include events with
semi--leptonically decaying heavy quarks; $W^+W^-$ events where one gauge boson
decays leptonically; and $ZZ$ events where one $Z$ decays into $\nu
\overline{\nu}$. Mono--jets might be a particularly important background to
one--sided
or ``Zen'' events \cite{39} that could signal the associate production
of a heavy and a light supersymmetric neutralino.

Here we consider mono--jets from two--photon events, as well as from $\ee
\rightarrow \qqbar$ annihilation events with hard photon emission from the
initial state, where again both beamstrahlung and bremsstrahlung are included.
In fig. 8a we show results for the two most extreme designs of \ee\ colliders
with \rs\ = 500 GeV listed in table 1, Palmer G and TESLA. We see again that
an increase of \Y\ increases the \gaga\ flux much more rapidly than the \ee\
flux at invariant mass well below \rs. At TESLA, two--photon events only
dominate for $\pT \leq 24$ GeV, while at Palmer G they continue to dominate
total mono--jet production up to $\pT \simeq 32$ GeV, and make important
contributions also in the region 45 GeV $\leq \pT \leq$ 55 GeV.

The spectrum of the annihilation contribution is, as usual, largely determined
by kinematic considerations. The cuts on the rapidities of the two jets
imply \be \label{e23}
\pT \leq \frac {\sqrt{s}}
{ e^{-|y_{1,{\rm max}}|} + e^{|y_{2,{\rm min}}|} } \simeq 0.131 \rs, \ene
where in the second step eqs.(\ref{e22}) have been used; this bound also
applies for two--photon events, of course. Fig. 8a shows that the
annihilation contribution stays at the level of 1 fb/GeV almost all the
way to the kinematical limit; recall that 1 fb corresponds to at least 10
events per year (up to 60 at Palmer G). Obviously the annihilation
contribution increases quite rapidly if the two jets can originate from
the decay of a real $Z$ boson.  For given transverse momentum, the
invariant mass of the \qqbar\ pair is minimized when $y_1 = y_{1,{\rm
max}}$ and $y_2 = y_{2, {\rm min}}$; on--shell $Z$ bosons can therefore
only contribute if \be \label{e24}
\pT \leq \frac {m_Z} { \sqrt{ 2 \left[ 1 + \cosh ( y_{2,{\rm min}} -
y_{1,{\rm max}} ) \right] } } \simeq 44.2 \ {\rm GeV}. \ene

However, the contribution of real $Z$ bosons will be suppressed if this final
state can only be produced by radiation off both electron legs. On--shell $Z$
bosons produced via single beam-- or bremsstrahlung only contribute if
\be \label{e25}
\pT = \frac {\sqrt{s}} { \frac {s} {m_Z^2} e^{-y_1} + e^{y_1} }.\ene
The r.h.s. reaches its absolute maximum of $m_Z/2$ at $y_1 = \log \rs/ m_Z$;
this, however, is in conflict with the constraint (\ref{e22a}) for the machines
we are considering. The maximal achievable \pT\ is therefore bounded by the
r.h.s. of eq.(\ref{e25}) with $y_1 = y_{1,{\rm max}}$. Of course, we also
have to require that $y_2 \geq y_{2,{\rm min}}$. On--shell $Z$ bosons produced
via the emission of a single photon from the initial state can therefore
only contribute if \be \label{e26}
\pT \leq \frac {\sqrt{s}} { \max ( \frac {s} {m_Z^2} e^{-y_{1,{\rm max}}},
 e^{y_{2,{\rm min}}}) + e^{y_{1,{\rm max}}} }. \ene
For \rs\ = 500 GeV, the r.h.s. amounts to 42.1 GeV; this is so close to the
value of eq.(\ref{e24}) that no extra structure at this point is visible
in fig. 8a. However, the near--equality of (\ref{e24}) and (\ref{e26})
explains why around the Jacobian peak of the $Z$ boson, the cross section
is actually smaller at Palmer G; in most real $Z$ events that pass the cuts
(\ref{e22}) at \rs = 500 GeV, the energy of one of the electrons is very
close to the nominal beam energy, where the flux at Palmer G is depleted due
to strong beamstrahlung, while the energy of one emitted photon is
so large that it is in most case produced by bremsstrahlung even at Palmer G.
Finally, we note that real $Z$ bosons can only be
produced in accordance with the cuts (\ref{e22}) if $y_1$ and $y_2$ have the
same sign. In contrast, the absolute upper bound (\ref{e23}) is saturated
if $|y_1 - y_2|$ is maximal, i.e. $y_1$ and $y_2$ have opposite signs. In
the region $\pT \geq 50$ GeV the cross section at Palmer G therefore
exceeds the one at TESLA again, since we are now in a region
where the fractional momenta of both the electron and the positron are
sizeable.

In fig. 8b we display the three classes of two--photon contributions
separately, for the case of the \gaga\ collider. Of course, there is no
\ee\ annihilation contribution here. Moreover, when computing the
kinematic limit (\ref{e23}),
\rs\ has to be replaced by $x_{\rm max} \rs \simeq 0.828 \rs$; the curves in
fig. 8b therefore terminate at a somewhat smaller value of \pT\ than those in
fig. 8a. We see that the spectrum shows an even steeper threshold at the
kinematical limit than do the annihilation contributions in fig. 8a; just 2 GeV
below the maximum, the direct contribution still amounts to 10 fb/GeV. This is
partly due to the slower decrease of the $\gaga \rightarrow \qqbar$ cross
section with increasing energy, compared to the $\ee \rightarrow \qqbar$
cross section. Furthermore, since the photon spectrum at the \gaga\ collider
is quite flat (see fig. 1a), configurations close to the edge of the phase
space region defined by the cuts (\ref{e22}) are not particularly
suppressed, unlike the situation at the TESLA collider.

Fig. 8b also shows that the once resolved contribution plays an important
role; we already saw in fig. 6b
that asymmetric cuts, like in (\ref{e22}), favour this contribution. However,
the most asymmetric initial state configuration ($y_1 = y_{1,{\rm max}}, y_2 =
y_{2,{\rm min}}$) only contributes to part of the \pT\ spectrum; for the
given case, it disappears for $\pT > 32$ GeV, which explains the small
kink that occurs in the 1--res spectrum at this point. In contrast, the
twice resolved contribution dominantly comes from rather symmetric initial
state configurations, which imply that $y_1$ and $y_2$ have opposite
signs, as can be seen from eqs.(\ref{e17}). Of course, $\pT \rightarrow
p_{T,{\rm max}}$ implies that $x \rightarrow 1$ for the parton densities
inside the photon; the 2--res spectrum in the threshold region is
therefore not as steep as the direct or 1--res spectrum. Finally, we
remark that the dependence of the relative importance of the three classes
of two--photon contributions at the \ee\ colliders is quite similar to the
case of di--jet production, discussed in some detail in the previous
subsection.

Fig. 8a shows that at \rs\ = 500 GeV, at least the hard part of the mono--jet
spectrum will be dominated by annihilation events, largely due to the
contribution from real $Z$ bosons. In Fig. 9 we compare the annihilation and
two--photon contributions for the three stages of the JLC. We see that already
at \rs\ = 1 TeV, the two--photon contribution dominates over almost the
whole kinematically accessible region. In particular, the contribution of real
$Z$ bosons now amounts to at most 5\% of the two--photon contribution. The
reason is that now the limit (\ref{e26}) gives $\pT \leq 32$ GeV. Most of the
true Jacobian peak of the $Z$ (which occurs at $\pT = m_Z/2$, of course) is
therefore only accessible after emission of two hard photons, and is therefore
strongly suppressed. Notice that we did not change the cuts (\ref{e22}) when
increasing the beam energy. In reality it might be necessary to allow for
smaller values of $y_{2,{\rm min}}$ at higher energy, since the coherent
production of \ee\ pairs rapidly increases \cite{23,36} with increasing \Y.
Even with these fixed cuts, we find a rate of about 1,000 mono--jet events
with $\pT \geq 100$ GeV per year at JLC2, and 300 events per year with
$\pT \geq 150$ GeV at JLC3. At lower values of \pT, the rate shoots up
very rapidly, due to the two--photon contribution; for instance, at JLC2 we
expect about 35,000 mono--jet events with $\pT \geq 50$ GeV per year. We
are therefore lead to the conclusion that missing \pT\ by itself will only
be useful as a signal for "new physics" if it amounts to at least 20\% of \rs.

There is yet another source of mono--jet events in the standard model: Three
jet annihilation events where two jets go in forward and backward direction,
respectively, while the third jet emerges at a large angle. The cross section,
integrated over $\pT \geq \ptmin$, for the dominant configuration where the
central jet stems from the gluon can be estimated as \be \label{e27}
\sigma(\ee \rightarrow \qqbar g) \simeq \sigma(\ee \rightarrow \qqbar)
\cdot \frac {1} {20} \cdot \frac {\alpha_s} {\pi} \cdot f(\ptmin), \ene
where we have ignored both beam-- and bremsstrahlung.
f describes the relative weight for configurations where the $q$ and $\bar q$
have an opening angle of at least $150^{\circ}$, while the angle between the
$g$ and the $q$ or $\bar q$ has to be at least $10^{\circ}$, as dictated by
the cuts (\ref{e22}); numerically, $f \simeq 10$ (1) for \ptmin\ = 0.1
(0.35) \rs. The additional factor 1/20 comes from the requirement that the $q$
and $\bar q$ be approximately aligned with the beam pipes. Notice that this
contribution extends to larger values of \pT\ than the contribution with only
two hard partons in the final state: \be \label{e28}
p_{T,{\rm max}}(\qqbar g) = \rs \frac {\sin \theta_{\rm max}} { 1 +
\sin \theta_{\rm max}} \simeq 0.21 \rs, \ene
where $\theta_{\rm max}$ is the maximal angle of the forward and backward jets;
the cut (\ref{e22b}) corresponds to $\theta_{\rm max} = 15.4^{\circ}$.
Numerically, eq.(\ref{e27}) gives approximately 30 (3) fb for \ptmin\ = 25
(90) GeV at \rs\ = 500 GeV. Comparison with figs. 8a and 9 shows that this
contribution will only be important at very large \pT. It does therefore not
change our previous conclusion about the relative importance of contributions
from annihilation and two--photon processes to mono--jet events.

\setcounter{footnote}{0}
\subsection*{3c. Heavy quark production}
In this subsection we study the production of $c,\ b,$ and $t$ quarks at
future \ee\ colliders. It should be clear from the
results of the previous two subsections that the total cross sections for
the production of \ccbar\ and \bbbar\ pairs at these colliders will be
dominated by two--photon contributions. The direct process as well as the
single resolved photon--gluon fusion process contribute with essentially the
same strength as in case of jet production from light quarks. On the
other hand, the twice resolved contribution is strongly suppressed here,
since none of the processes that proceed via gluon exchange in the $t$ or
$u$ channel can contribute. We therefore expect this latter class of
contributions to be relatively less important here.

This is born out by the results of tables 2 and 3, where we list predictions
for total \ccbar\ and \bbbar\ cross sections; all contributing processes are
shown separately. As described in more detail in ref.\cite{6}, we have assumed
different values for the ``dynamical'' quark mass entering the matrix elements,
and the ``kinematical'' mass which determines the phase space. For charm and
beauty production we have used \cite{40} dynamical masses of 1.35 GeV and 4.5
GeV, respectively; the kinematical mass is always taken to be the mass of the
lightest meson carrying the corresponding heavy flavor. We do not list results
for the nbb version of the DESY-Darmstadt design, since they differ by only
10\% or less from those of the TESLA design; this difference is smaller than
the theoretical error of our esitmates.

We see that the direct \ccbar\ cross section varies much less between the
different designs than the resolved photon contributions do. This is
because, as shown in sec. 2, designs with smaller beamstrahlung parameter
\Y\ tend to have more soft photons, which contribute strongly to direct
\ccbar\ production, but have little impact on the resolved photon
contributions; for instance, the direct \ccbar\ cross section at TESLA is
about the same as at Palmer F, but the latter has an about 2.5 times
larger resolved photon contribution. Notice also that the \gaga\ collider
with its hard photon spectrum actually has the {\em smallest} direct
\ccbar\ and \bbbar\ cross sections; due to the huge 1--res contribution,
it nevertheless has by far the largest total \ccbar\ and \bbbar\ cross
sections.

The 1--res and direct contributions are of similar size
at the 500 GeV \ee\ colliders,
with the exception of the Palmer G design. At higher energies, however, the
resolved photon contributions clearly begin to dominate. Notice also that
the DG parametrization, which we used here, predicts
the ratio of 1--res and direct contributions to be roughly the same for \ccbar\
and \bbbar\ production; the more rapid decrease of the resolved photon
contribution with increasing mass is balanced by the charge suppression factor
of 1/4 of the direct contribution. The exception is the TESLA (and
DESY--Darmstadt (nbb)) design, where the
increase in mass also suppresses the direct contribution strongly, due to the
very soft beamstrahlung spectrum.

Tables 2 and 3 also contain an entry for the production of the 1s vector
quarkonium state. In leading order in $\alpha_s$, this state can only be
produced \cite{6} in resolved photon reactions; by far the dominant
contribution comes from the single resolved process $\gamma + g \rightarrow
J/\psi + g$, and correspondingly for the \Y(1s). We estimate these cross
sections using the color singlet model \cite{41}. The cross section for
$J/\psi$ production is so large that it should be easily detectable via its
decay into muons or electrons even at the TESLA collider. The cross sections
for \Y(1s) production are smaller by a factor of about 500; in addition,
the branching ratios for the leptonic decays are almost 3 times smaller than
for the $J/\psi$. Nevertheless, at least 15 \Y(1s) $\rightarrow \mu^+ \mu^-$
per year are expected to occur even at TESLA.

The results of tables 2 and 3 have been obtained using the conservative DG
parametrization for \qvec. The other parametrizations discussed in sec. 2
lead to larger predictions for the resolved photon contributions. We can
conclude from fig. 5a that the LAC2 parametrization predicts almost 5 times
larger 1--res \ccbar\ cross sections than DG does; in that case resolved
photon contributions would dominate \ccbar\ production at all colliders we have
studied here. However, even though the 2--res contribution would increase by a
factor of 15 or so, it would still be subdominant.
Since the LAC2 parametrization contains a very steeply falling \Gg$(x)$, it
predicts the ratio of resolved to direct contributions to decrease by
approximately a factor of two when going from \ccbar\ to \bbbar\ production.
The predictions of the modified DO+VMD parametrization for the single resolved
contribution also lie a factor 1.5 to 2 above those of the DG parametrization.

So far we have only discussed total cross sections. The results seem to
indicate enormous event rates, especially for \ccbar\ pairs. This might be
somewhat misleading, however, since in many events the heavy quarks emerge
at such small angles that they remain unobserved. In particular, we saw
in fig. 6 that single resolved \qqbar\ production is concentrated at large
rapidities, due to the asymmetric initial state. A realistic estimate of
the number of {\em observed} \ccbar\ and \bbbar\ events needs a full
simulation of the detector, which is beyond the scope of this paper.
One might be able to get an idea of the result of such a full simulation by
looking at the \pT\ spectrum of centrally produced heavy quark pairs. In fig.
10a,b we therefore show the transverse momentum spectrum of charm quarks
produced at $\theta = 90^{\circ}$, i.e. $y_1 = y_2 = 0$, for the TESLA (9a)
and Palmer G (9b) designs. We see that at TESLA the resolved photon
contribution is now well below the direct one for all \pT, while at Palmer
G it exceeds the direct contribution by at most a factor of two. Notice
that the relative importance of the 2--res contributions is actually
enhanced by going to small rapidities; due to the symmetric initial state
configuration and the soft parton distributions inside the electron the
twice resolved contribution is concentrated at small $y$. At TESLA, the
effective quark density in the electron is even softer than the gluon
distribution; we are again seeing the contribution from quarks with large
Bjorken--$x$ inside soft photons. Finally, fig. 10 shows that one can
neglect\footnote{The production of high--\pT\ charm quarks via resolved
photon mechanisms is probably dominated by flavor excitation processes,
rather than the pair production process we have studied here. If $p_T^2
\gg m_c^2$, one can again treat the charm quark as an essentially massless
parton inside the photon. However, the flavour structure of the photon is
not well understood \cite{31b}; no existing parametrization treats the
quark mass effects properly. In any case, although this process might be
interesting in itself, it will be subdominant at large transverse
momentum.} all resolved photon contributions if one is only interested in
central events with hard muons or electrons; such events might be
\cite{42} a background to top production.

It has recently been pointed out \cite{42} that \ttbar\ production at
future \ee\ colliders might itself be dominated by two--photon events. In
fig. 11a,b we compare \ttbar\ production via \gaga\ fusion and \ee\
annihilation at two designs of \ee\ colliders operating at \rs\ = 500 GeV
(10a), as well as the third stage of the JLC (10b). Notice that the
two--photon contributions in fig. 11a have been multiplied with 10 (for
Palmer G) and 100 (for TESLA), respectively. We see that at \rs\ = 500
GeV, \gaga\ processes will contribute at most 7\% of the total \ttbar\
cross section; their contribution at TESLA is always well below 1\%. In
fact, it might be very difficult to even detect the two--photon
contribution, since some annihilation events will also have a \ttbar\
invariant mass well below \rs, due to the combined effects of brems-- and
beamstrahlung (see fig. 2).

Fig. 11b shows that at \rs\ = 1.5 TeV the two--photon contribution could indeed
dominate, but not by a large factor; moreover, for $m_t \geq 125$ GeV the
annihilation contribution is still the more important one. In this figure the
direct and total \gaga\ contributions are shown separately; even though we have
used the modified DO+VMD parametrization with its hard (and $Q^2$ independent)
intrinsic gluon component here, we still find that at this collider, at most
10\% of the total two--photon contribution comes from resolved photons.
They can make important contributions to \ttbar\ production only at \gaga\
colliders operating at $\rs \geq 1$ TeV.

Figs. 11 also show that an estimate of the annihilation contribution to
\ttbar\ production should include the effects of beam-- and bremsstrahlung.
At \rs\ = 500 GeV, they increase (decrease) the cross section for
$m_t < (>)$ 155 GeV. For light top quarks, the increase of the photon and
$Z$ propagators $\propto 1/m^2_{t \bar t}$ is the dominant effect, while for
large $m_t$, the reduction of the available phase space is more important.
At \rs\ = 1.5 TeV, the top quark is always ``light'', of course. Radiation
therefore increases the cross section by a factor between 1.5 and 1.7; it also
leads to a decrease of the annihilation cross section by about 10\% when $m_t$
is increased from 90 to 200 GeV. About 30 to 40\% of these effects is due to
bremsstrahlung; beamstrahlung by itself increases the annihilation cross
section by about 30\% at JLC3. While this is certainly not negligible, it pales
compared to the 800\% increase of the \gaga\ contribution to \ttbar\ production
which is also caused by beamstrahlung at this collider. Nevertheless, the total
\ttbar\ cross section at \rs\ = 1.5 TeV remains considerably smaller than at
\rs\ = 500 GeV.

\setcounter{footnote}{0}
\subsection*{3d. Single $W$ and $Z$ production}
We now turn to our final example of a hard two--photon process, the production
of a single $W$ or $Z$ boson. The corresponding processes at $ep$ colliders
like HERA have been studied in some detail in the literature \cite{44}. In
particular, it has been shown that the total cross section can to 20--30\%
accuracy be estimated from the simple resolved photon process \cite{45} $\qqbar
\rightarrow W,Z$. We will assume that this is also true for two--photon
reactions, and will estimate the total cross sections from the twice
resolved \qqbar\ annihilation (Drell--Yan) process alone.

Our results are summarized in table 4, where we list the total cross sections
for single $W$ and $Z$ production in two--photon collisions
at various colliders.
The $W$ cross section includes $W^+$ as well as $W^-$ production; since the
initial state has even $C$ parity, the $W^+$ and $W^-$ cross sections are, of
course, equal. The results of table 4 have been obtained using the DG
parametrization with $Q^2 = m^2_{W,Z}$. Increasing $Q^2$ by a factor of two
increases the cross sections by about 10--20\%; note that in the given case
the increase of \qvec\xq\ and $f^{\rm brems}_{\gamma|e}$ is not compensated
by a decrease of the hard cross section, in contrast to the reactions we
studied in secs. 3a--c. The modified DO+VMD parametrization predicts 50--70\%
larger $W$ cross sections, and 30--50\% larger $Z$ cross sections, where the
larger number refers to the hardest photon spectra (JLC3 and the \gaga\
colliders). However, at least part of this excess is certainly fake. As
noted before, the VMD contribution is assumed to be $Q^2$--independent,
which overestimates its importance at high $Q^2$. Furthermore, the
DO parametrization only includes $N_f = 4$ active flavours, while for
$Q^2 \simeq m_W^2, \ N_f = 5$ seems more appropriate. The $b$ quark itself
does not contribute to $W$ production, but the increase of $\alpha_s$ when
going from 4 to 5 flavours leads to somewhat softer quark distributions
in the photon.

This is also one of the few processes where the flavour structure of the
photon plays an important role.  For example, for the LAC2 parametrization
we find $W$ cross sections larger by 70--150\% and $Z$ cross sections higher by
about 30--100\% at the JLC3 and $\gamma \gamma $ colliders. While it is
true that this parametrization again uses only $N_f = 4 $  and
part of the increase may be ascribed to that, the real reason for this
difference lies in the different flavour structure of the DG and LAC2
parametrization. The LAC2 parametrization does not satisfy the constraint
$u^\gamma(x) = 4 d^\gamma(x) $ even at large $x$ and $Q^2$. As a
result it requires  a higher $d$ quark content of the photon
(as compared to the DG
parametrization) to  fit the data on $F_2^\gamma $. This leads to
higher cross sections for both $W$ and $Z$ production.

By comparing the results of table 4 with the integrated di--jet cross sections
shown in figs. 3 and 7 one can immediately convince oneself that it will be
very difficult to observe the gauge bosons in their hadronic decay modes.
(This is again similar to the case of HERA \cite{46}.) One will thus have to
use leptonic decay modes. $W$ production would therefore be signalled by a
hard lepton, with a Jacobian peak at $m_W/2$ in its \pT\ spectrum, in
association with large missing transverse momentum; the signal for $Z$
production is simply a hard lepton pair whose invariant mass equals $m_Z$. In
both cases the event should contain two spectator jets. We remind the reader
that the branching ratio for the leptonic decays are only 11\% and 3.3\%
per generation for the $W$ and $Z$ boson, respectively. Even after summing
over $e$ and $\mu$ channels, we therefore only expect about 10 (35)
detectable $W$ events per year at TESLA (JLC1). This should be compared to an
$\ee \rightarrow W^+ W^-$ cross section of about 8 pb; the cross sections
for the annihilation processes $\ee \rightarrow W e \nu$ and $\ee \rightarrow
\ee Z$ also amount to about 5 pb at \rs\ = 500 GeV \cite{47} even if
beamstrahlung can be ignored. Of course, these annihilation events lack the
spectator jets of the resolved photon events; moreover, the gauge bosons are
usually produced with sizeable transverse momentum. Nevertheless, it is
quite clear that extraction of the two--photon signal will be quite
difficult, if not impossible, at a 500 GeV \ee\ collider.

The situation might be different, however, at higher energies. At the second
stage of the JLC, we expect as many as 1500 $W \rightarrow l \nu$ and 230
$Z \rightarrow l^+ l^-$ events from two--photon processes per year ($l = e,
\mu$). The cross sections for the single production of a gauge boson also
increase when going from \rs\ = 500 GeV to 1 TeV, but only by about a
factor of two \cite{47}\footnote{However, the calculation of Hagiwara et
al. \cite{47} does not include the contribution from beamstrahlung photons,
e.g. $\gamma e \rightarrow W \nu$, which should be quite large at this
collider.}. The rates at \gaga\ colliders are even larger; assuming an
integrated luminosity of 20 $fb^{-1}$ per year, one has about 1,750
$W \rightarrow l \nu$ events and 275 $Z \rightarrow l^+ l^-$ events per
year already at \rs\ = 500 GeV. Notice, however, that the $\gaga \rightarrow
W^+ W^-$ cross section amounts to about 80 pb, giving as many as 500,000
events with one leptonically decaying $W$ boson per year; extraction of
the $W$ signal will therefore still not be trivial. On the other hand, the
background for the $Z$ signal should be much smaller. Finally, we remark that
at \gaga\ colliders, our cross sections increase almost linearly with
energy, while the $\gaga \rightarrow W^+ W^-$ cross section stays constant.

\section*{4. Semi-hard and soft two-photon reactions}
\setcounter{footnote}{0}
In this section we discuss semi--hard (minijet) and soft (VMD) two--photon
reactions at future \ee\ linacs \cite{10}. ``Semi--hard'' here merely means
that we are trying to push leading order perturbative QCD to its limit of
applicability. We do not attempt to re--sum $\log 1/x$ terms, or to include
shadowing effects. The main emphasis will be on the question whether these
events give rise to an ``underlying event'', where one or several
two--photon reaction occurs simultaneously (within the time resolution of
the detector) with every annihilation event; if such an underlying event
does occur, we try to characterize it at least qualitatively.

We have already seen in figs. 3 that the cross section for the production
of a pair of jets in two--photon collisions increases very rapidly with
decreasing transverse momentum of the jets; figs. 3b and 4 show that this is
mostly due to the contribution from resolved photon processes. In figs.
12a,b we extend these calculations to even lower values of the minimal
transverse momentum \ptmin\ of the partons participating in the
(semi--)hard $2 \rightarrow 2 $ scattering process. We show results for the
DG (12a) and modified DO+VMD (12b) parametrization; since we are now
considering reactions that are characterized by a relatively low momentum
or $Q^2$ scale, the effect of the $Q^2$ variation of the hadronic VMD
contribution to \qvec, which we ignored, is probably not very large here.
Notice that we have not applied any rapidity cuts in figs. 12. Due to the
nontrivial colour flow between spectator and ``hard'' jets, a resolved
photon event should always include some detectable particles, even if the
``hard'' jets emerge at very small angles, and will thus always contribute
to the underlying event. A direct event might remain invisible if both jets
are produced in the very forward or very backward direction, due to a strong
boost between the \gaga\ centre--of--mass frame and the lab frame; however,
less than 1\% of all minijet events will come from the direct process at the
colliders we are considering.

Unfortunately, figs. 12 show that the leading order prediction for the cross
section depends quite sensitively on \ptmin. This is not surprising, since
most of the hard $2 \rightarrow 2$ cross sections diverge like
$1/p_{T, {\rm min}}^2$ as $\ptmin \rightarrow 0$. An additional \ptmin\
dependence is produced by the growth of the parton densities at low $x$.
Eqs.(\ref{e17}) and (\ref{e18}) show that the average $x$ decreases linearly
with \pT, while the kinematical minimum of $x$ even decreases quadratically
with decreasing \pT. The results of figs. 12 can to good approximation be
parametrized by a power law, $\sigma (\pT \geq \ptmin) = a
p_{T, {\rm min}}^{-b}$, where the power $b$ is approximately independent
of the photon spectrum (i.e., of the collider), but does depend on
the parametrization we used; one has $b \simeq 3.3$ (3.6) for the DG
(modified DO+VMD) parametrization. The prediction for the cross section
therefore changes by a factor of 2 when \ptmin\ is changed by 23 (21) \%!
It is therefore very important to at least try to get an idea down to which
value of \ptmin\ our calculation might be reliable.

We see from eq.(\ref{e19}) that \ptmin\ determines the minimal virtuality
of the exchanged parton in the $2 \rightarrow 2$ scattering, and thus the
``hardness'' of the process. It should therfore be analogous to the momentum
transfer $Q^2$ in deep inelastic scattering. Standard parametrizations of
hadronic structure functions \cite{48}, which rely on the validity of
perturbative QCD, are assumed to be reliable down to some value $Q_0^2$
in the range between 1 and 5 $GeV^2$. Further support for the applicability
of perturbative QCD at momentum scales between 1 and 2 GeV comes from its
success in describing at least the gross features of charmonium physics
\cite{49}, as well as of open charm production from hadrons \cite{50}.

Moreover, minijet calculations are also able to reproduce quite well
the observed rise of the total \ppbar\ cross section with energy. The basic
idea that semi--hard QCD interactions could affect such a seemingly ``soft''
quantity as the total cross section dates back to 1973 \cite{51}. Of course,
minijet calculations for \ppbar\ reactions also depend on a cut--off \ptmin.
Recent fits to existing data \cite{52}
indicate that \ptmin\ has to be chosen in the range
between 1.3 and 2 GeV, if the rise of hadronic cross sections is to be
described by minijets. It is sometimes even claimed that minijets have been
seen experimentally by the UA1 collaboration \cite{53}. However, the UA1
analysis only inlcuded ``clusters'' with transverge energy of at least 5 GeV,
which corresponds to a minimal partonic \pT\ of approximately 3.5 GeV. The
cross section for the production of such clusters does indeed grow very
rapidly with energy, in the region 200 GeV $\leq \rs \leq$ 900 GeV, in
accordance with leading order QCD predictions. However, we have seen above
that changing \ptmin\ from 3.5 to 1.5 GeV would change the leading order
prediction of the cross section by more than an order of magnitude. In our
opinion the UA1 results are therefore not a direct proof for the validity of
the minijet ansatz, although they are certainly not in disagreement with it.
In fact, it seems quite unlikely that ``jets'' with (partonic) \pT\ as small
as 1.5 to 2 GeV can ever be identified at hadron colliders.

Fortunately the situation is quite different for two--photon collisions, where
``jets'' with \pT\ as small as 1 GeV are routinely reconstructed \cite{35}.
The relationship between the transverse momenta of the parton and the
resulting jet is quite complicated, however. At such small values of \pT,
contributions from the hadronization process, as well as from the intrinsic
\pT\ of the partons, are not negligible. Moreover, the whole event is forced
into a two--jet topology; parts of the spectator jets of resolved photon
events will thus be included in the reconstructed jets. One therefore needs
a careful Monte Carlo analysis to derive the partonic \pT\ from the transverse
momentum of the jets even on a statistical basis. So far the only analysis
of this kind has been performed by the AMY collaboration \cite{7}, using their
data taken at the TRISTAN collider at $\rs \simeq 60$ GeV. Their Monte Carlo
generator was able to describe the real data quite well, both in shape and
normalization, once the resolved photon contributions had been taken into
account; this is in sharp contrast to older analyses \cite{35} where these
contributions were not included, and consequently an excess of data over the
Monte Carlo predictions was oberved. AMY determined the minimal partonic
\pT\ using only events with $\pT (jet) \geq 3$ GeV, where the soft or VMD
component, which is characterized by an exponential \pT\ spectrum, is already
essentially negligible; they found \ptmin\ = 1.6 (2.4) GeV for the DG
(DO+VMD) parametrization. These numbers depend only weakly on the chosen
fragmentation and hadronization scheme. At least in case of the DG
parametrization, the AMY value for \ptmin\ falls within the range of values
favoured by other minijet analyses. We will therefore from now on use their
values as out best guess for \ptmin.

The resulting predictions for the total semi--hard two--photon induced
cross section at a variety of hypothetical future colliders are listed in
columns 2 and 3 of table 5. We see immediately that the modified DO+VMD
parametrization predicts a 1.4 to 1.7 times smaller cross section than the
DG parametrization; the increase of \Gg\ is over--compensated by the increase
in \ptmin. However, we should caution the reader that this is partly due to
our rather arbitrary regularization (\ref{e10}) of the original DO
parametrization \cite{14}. Without this regularization, this prediction
would be approximately 20\% higher at the 500 GeV \ee\ colliders; the
effect of the regularization is even larger for harder photon spectra and
higher electron beam energies.

In order to translate the cross sections of table 5 into a meaningful number
of events, we first define an ``effective bunch crossing''. If within the
time resolution of the detector only one bunch crossing occurs, the
luminosity per effective bunch crossing is identical to the luminosity per
bunch crossing \lhat\ listed in table 1. If the temporal separation of
consecutive bunches $\Delta t$ is smaller than the time resolution $\delta t$,
we sum over $\delta t / \Delta t$ bunch collisions, or over a complete bunch
train collision, whatever gives the smaller number. (Consecutive bunch train
collisions can trivially be distinguished.) In table 5 we have assumed a
rather poor time resolution of $10^{-7}$ seconds, which should be quite easy
to achieve. However, table 1 shows that only the DESY--Darmstadt design
would benefit from an improved time resolution of $3 \cdot 10^{-8}$ sec. In
any case it is trivial to compute the effects of better time resolution from
the numbers in the last column of table 5.

These numbers indicate that most designs for \ee\ colliders operating at
\rs\ = 500 GeV should have at most one event per average effective
bunch collision. Since these events should obey a Poisson distribution,
an average of one event per effective bunch crossing still means that more
than 35\% of all bunch collisions are free of any two--photon events,
independent of whether they contain an annihilation event or not. This would
be equivalent to a reduction of the luminosity by a factor of 3 for those
measurements where not even a single
two--photon event can be tolerated, {\em if}
the presence of a two--photon event can be reliably detected when
an annihilation event occurs at the same time. For instance, the measurement
of the mass of the top quark to sub--GeV precision \cite{54} is not limited
by statistics; such a measurement could thus still be performed at the JLC1
collider, if \ttbar\ events that also contain a two--photon
event can be reliably distinguished from ``pure'' \ttbar\ events.

On the other hand, performing such a measurement at the Palmer G collider
will be almost impossible, if our estimate for the number of two--photon
events that occur at each effective bunch crossing is at least approximately
correct. Assuming that the spectator jets deposit about 1 -- 2 GeV
transverse energy per unit of rapidity, and adding another 4 GeV if the
``hard'' jets are produced centrally, we estimate that each minijet event
will deposit between 5 and 12 GeV transverse energy in the central part of
the detector, defined by the rapidity window $|y| \leq 2$. At Palmer G one
would therefore have to expect at least 100 GeV of transverse energy in soft
particles to underly every annihilation event; a similar number has to be
expected at the second stage of the JLC, and the third stage would be even
worse. This would cause a host of problem familiar from hadron colliders.
Examples are: a deterioration of the experimental resolution of jet energies,
which would, e.g., make it difficult to distinguish between hadronically
decaying $W$ and $Z$ bosons; a large number of tracks,
which complicates $b$--tagging with microvertex detectors; fluctuations in
the underlying event, which could produce missing tranverse momentum; and
difficulties in defining isolation criteria for hard leptons, which
figure prominently in searches for semi--leptonically decaying heavy
particles.

The fourth column of table 5 shows the VMD prediction for the total
hadronic cross section for events with \gaga\ invariant mass $\wgg \geq 5$
GeV, assuming a constant $\gaga \rightarrow hadrons$
cross section of 250 nb \cite{55}. We see that
for the 500 GeV \ee\ colliders the predicted minijet cross section always
falls below this conservative estimate of the total \gaga\ cross section.
In principle, the contributions from both these sources should be included, if
one wants to estimate the total number of events; for instance, the AMY
Monte Carlo generator needs both soft and hard two--photon reactions to
explain their data. However, it is not clear whether a soft interaction
will always be observable at high energy \ee\ or \gaga\ colliders. At
low \wgg, the multiplicity of soft events seems to be quite low, at least
according to standard MC generators \cite{56}. No experimental information
exists about two--photon events
with $\wgg > 25$ GeV or so, but it seems possible
that (part of) the soft component becomes diffractive, so that (almost) all
particles are concentrated in the forward and backward regions. In any case,
it is quite certain that the average $E_T$ in a soft event will be smaller
than in a minijet event. We will also see below that it may no longer be
appropriate to simply sum the soft and hard contributions to the total \gaga\
cross section if the hard contribution is of the same order as or larger than
the soft one. For these reasons we have ignored the soft contribution when
estimating the number of events per effective bunch crossing.

The results for the second and third stage of the JLC show that building a
``clean'' \ee\ collider with $\rs \geq 1$ TeV might be quite difficult. The
same conclusion also holds for simple extrapolations of the $X$--band
design with the smallest minijet cross section, Palmer F. In principle it
might be possible to improve the time resolution of the detecor to something
like 2 nanoseconds; the drift velocity of electrons in gas seems to make it
impossible to achieve better time resolution with present technology
\cite{57}.
Even a time resolution of 2 nanoseconds seems quite difficult to achieve,
given that an ultrarelativistic particle needs about 10 to 15 nanoseconds to
traverse the detector; at the JLC bunch spacing of 1.4 nanoseconds fast
particles produced in the current bunch crossing can therefore overtake
slower particles produced in previous bunch crossings. The problem is
further complicated by the probable occurence of ``loopers'', i.e. of
particles describing spiral orbits in the magnetic field of the detector,
which could stay in the detector for several $10^{-8}$ seconds. The assignement
of a given particle to a certain bunch crossing can therefore only occur
on the software level, by combining information about arrival times and
energy/momentum of the particle, or by reconstructing its track. In view of
these problems it seems unlikely that a detector for an \ee\ supercollider
would be much easier to build than a detector for a $pp$ supercollider,
{\em if} such a time resolution turns out to be necessary. Note also that
even with this excellent resolution, the leading order DG calculation
still predicts 2.5 (5) two--photon events to occur per effective bunch
collision at the JLC2 (JLC3).

Nevertheless it might be possible to build TeV linear colliders with $\ll 1$
events per effective bunch crossing. This is demonstrated by the last 4 rows
of table 5, where we have tried to extend the TESLA design as described in
table 1 to higher energies. As discussed in sec. 2, quite simple considerations
show that the beamstrahlung parameter \Y\ should grow between linearly and
quadratically with the beam energy; in the first case one assumes a constant
luminosity per bunch crossing \lhat, while in the second case \lhat\ grows like
$s$. Our predictions for the first, more optimistic scenario are given in
rows 9 and 10, while rows 11 and 12 show results for the less favorable
extrapolation; in both cases the number given in the first column is \rs\ in
GeV. We don't claim these to be realistic extrapolations; e.g., we have not
varied the bunch length at all. Nevertheless, they should be sufficient to
give us some indication of the true situation.

We see that in the optimistic scenario one can achieve a clean environment
even at \rs\ = 2 TeV, with a large safety margin. Of course, the total
luminosity has to grow like $s$ if the machine is to retain its full
potential. Since we assumed constant luminosity per bunch crossing, one has
to increase either the number of bunches per train, or the number of bunch
train collisions per second. At worst, one would have to increase the number
of bunches by a factor of 16 when going from \rs\ = 0.5 TeV to 2 TeV; this
would still leave a time gap between subsequent bunches of 60 nanoseconds, so
that single bunch collisions could be resolved quite easily. Of course, the
large safety margin shows that one might for technical reasons prefer to
increase \lhat\ at least slightly, even at the cost of a somewhat more
rapid increase of \Y. However, our results for the less favourable projection
show that even at a TESLA--like design one would have to deal with an
underlying event at $\rs > 1.5$ TeV, if both \Y\ and \lhat\ grow quadratically
with the beam energy. Recall that they grow between linearly and quadratically
at the JLC collider as currently planned.

Finally, notice the very large minijet cross section at the \gaga\ collider
already at \rs\ = 500 GeV. If the \gaga\ collider originates from an \ee\
collider like TESLA, one still only expects one event every 2 bunch crossing or
so; a similar rate can be achieved at the DESY--Darmstadt (nbb) design, if a
time resolution of around 50 nanoseconds can be realized.
At all the other designs one would have to expect (much) more than one
event per effective bunch collision; the higher number given in table 5
corresponds to the Palmer G design. In all cases the minijet rate at a \gaga\
collider would be far larger than at its \ee\ progenitor. The \gaga\ option,
while interesting in its own right \cite{58}, would therefore not help to
solve the problem of hadronic backgrounds.

Note that the leading order prediction for the minijet cross section at the
\gaga\ collider is far {\em above} the VMD prediction for the {\em total}
hadronic cross section; clearly at least one of these predictions must be
wrong. The problem is illustrated in fig. 13, where we show the DG prediction
for the total minijet cross section (with \ptmin\ = 1.6 GeV) as a function of
the \gaga\ centre--of--mass energy \wgg. It obviously rises very quickly with
energy. The AMY analysis \cite{7} provided experimental evidence for the rapid
growth of the resolved photon cross section when going from PETRA to
TRISTAN energies, but this only probes the region $\wgg \leq 25$ GeV.
We find that the leading order
prediction exceeds the VMD prediction of 250 nb for $\wgg \geq 50$ GeV. The
true value of the soft two--photon cross section even at low energies is
quite uncertain; even a number as large as 420 nb has been quoted \cite{59}.
Moreover, one might envision a slow (logarithmic) increase of the soft
cross section with energy. On the other hand, the result of fig. 13 can be
parametrized as \be \label{e29}
\sigma^{LO}({\rm DG}) = 250 \ {\rm nb} \left( \frac {W_{\gamma \gamma}}
{ 50 \ {\rm GeV}} \right)^{1.4} ; \ene
this reproduces the numerical leading order prediction to better than 10\% in
the region 10 GeV $\leq \wgg \leq$ 500 GeV. This cross section will be
substantially larger than any conceivable VMD estimate in the region $\wgg
\geq 100$ GeV.

Fig. 1 shows that the photon luminosity at \ee\ colliders decreases quite
rapidly with \wgg; on the other hand the leading order cross section
(\ref{e29}) clearly gives great weight to the region of large \wgg.
It is therefore possible that the region of \wgg\ where $\sigma^{\rm LO}
({\rm DG}) > \sigma$(VMD) contributes significantly to the total minijet
cross section at a given collider, even if that cross section is still
below the total VMD cross section at the same collider. This is demonstrated
by the dashed curves in fig. 13, which refer to the scale on the right side
of the frame; they depict the fraction of the total minijet cross section at
a given collider that comes from events with two--photon invariant mass
smaller than the \wgg\ shown as the $x$-axis. We see that for the Palmer F
design, most minijet events still have values of \wgg\ where the leading
order prediction for the hard cross section is less than or roughly equal to
the VMD prediction for the total cross section; the same is true for the
TESLA, DESY--Darmstadt and JLC1 designs. On the other hand, at the Palmer G
collider 30\% of the minijet events have $\wgg > 100$ GeV, where the DG
prediction (29) clearly exceeds the VMD estimate. This feature becomes even
more prominent for harder photon spectra, as shown by the curves for the
JLC2 and \gaga(500) colliders.

The problem that the leading order prediction for the minijet cross
section exceeds the total cross section if \ptmin\ is chosen in the GeV
range is well known in the case of hadronic collisions. The standard
remedy \cite{52} is to interpret the leading order calculation not as a
prediction of the total cross section $\sigma$, but as a prediction of
$\sigma$ times the number of minijet pairs per event. In this way two
events with one jet pair each can be combined into one event with two
pairs of jets. Formally, this is achieved by eikonalizing the cross section.
Essentially one writes \cite{52} \be \label{e30}
\sigma_{p \bar p}^{\rm inel} = \int d^2 b \left[ 1 - e^{- \left( \pphard
(s) + \csoft \right) A(b) } \right] . \ene
$A(b)$ describes the transverse distribution of partons in the proton,
normalized such that $\int d^2 b A(b) = 1$. \csoft\ is assumed to be (almost)
independent of $s$. It is related to the soft inelastic \ppbar\ cross section,
which is essentially equal to the total inelastic \ppbar\ cross section at
low energies, by \be \label{e31}
\ppsoft = \int d^2 b \left[ 1 - e^{-\csoft A(b)} \right] . \ene
If $\csoft \ll 1/A(0)$\footnote{Obviously, $A(b)$ is maximal at $b=0$.}, we
simply have \csoft =\ppsoft; moreover, eq.(\ref{e30}) reduces to
$\sigma_{p \bar p}^{\rm inel} = \pphard + \ppsoft$ if the hard cross section
is also small in this sense. However, if either cross section is large, of
order of the geometrical cross section, eq.(\ref{e30}) predicts a much
slower increase of the total cross section with energy than predicted by the
simple leading order calculation.

Unfortunately, it is not entirely straightforward to apply this formalism
to reactions involving photons in the initial state. As first pointed out
by Collins and Ladinsky \cite{60} for the case of the $\gamma p$ cross
section, the ansatz (\ref{e30}) has to be modified. The point is that
once a photon has ``transformed'' itself into a (virtual, but long--lived)
hadronic state, which it does with probability $\ph \simeq {\cal O} (\aem)$,
the production of additional pairs of jets should {\em not} be suppressed
\cite{61} by additional powers of \aem, as would be predicted by eq.(\ref{e30})
since \gghard\ is itself ${\cal O} (\alpha_{\rm em}^2)$. For the case of
\gaga\ collisions one has to write \cite{storrow} \be \label{e32}
\sigma_{\gamma \gamma}^{\rm inel} = \int d^2 b P_{\rm had}^2 \left[ 1 -
e^{- \left( \gghard(s) + \ggsoft \right) A(b) / P_{\rm had}^2 } \right]. \ene
The problem is that it is not at all obvious how $A(b)$ and \ph\ are to be
determined. For instance, it is generally accepted that \ph\ should be of
order \aem, but it is not clear just how large it is. From the VMD model,
one estimates \cite{60} $\ph \simeq 1/300$; in this case eikonalization
reduces the minijet cross section at the 1 TeV collider of ref.\cite{11} by
approximately a factor of 2 \cite{storrow}. On the other hand,
parton model considerations
lead to the estimate \cite{61} $\ph \simeq 1/170$. Recently it has been
suggested \cite{62} that \ph\ might even grow logarithmically with energy,
so that \ph(100 GeV)$\simeq 1/55$. Obviously the asymptotic hadronic
\gaga\ cross section as predicted by eq.(\ref{e32}) is proportional to
$P^2_{\rm had}$; even if \ptmin, the parton densities inside the photon and
$A(b)$ were all known, estimates of $\sigma_{\gamma \gamma}$ would still
differ by a factor of 30! In particular, it is very well possible that even
after eikonalization the cross section exceeds the VMD estimate substantially.
In fact, one can argue that the smallness of the VMD cross section (250 nb)
is hard to understand from perturbative QCD. Once hard interactions start
to dominate the exponents in eqs.(\ref{e30}) and (\ref{e32}), one would
expect the \gaga\ cross section to lie very roughly between $\alpha_{\rm em}^2
\sigma_{p \bar p}$ and $(\aem/\alpha_s(\ptmin))^2 \sigma_{p \bar p}$,
i.e. between
2 and 25 $\mu$b for \wgg\ = 500 GeV; recall that $\qvec \propto 1/\alpha_s$.
If this simple estimate is at least halfway correct, the total hadronic
\gaga\ cross section could be described by eq.(\ref{e29}) for $\wgg \leq 200$
GeV, and possibly up to $\wgg \simeq 1$ TeV. Fortunately, in the near future
measurements of the total $\gamma p$ cross section at centre--of--mass
energies up to about 250 GeV will be performed at HERA. Different ans\"atze
for \ph\ and $A(b)$ also lead to quite different predictions \cite{60,61,62}
for $\sigma_{\gamma p}$, so that one should be able to reduce the uncertainty
of theoretical predictions for $\sigma_{\gamma \gamma}$ by fitting model
parameters to those HERA data.

Finally, we would like to argue that, while the total \gaga\ cross section is
certainly of great theoretical interest, since it could teach us important
lessons about semi--hard QCD, it is in many cases {\em not} a good measure
for the severity of problems caused by soft and semi--hard two--photon
backgrounds. We see from table 5 that, whenever the leading order estimate
predicts $\gg 1$ minijet events per effective bunch crossing, so does the
conservative VMD estimate. Since it would be implausible to assume
that the total \gaga\ cross section at high energies is even smaller than the
VMD estimate, we can in such cases be sure that $\geq 1$ two--photon event
will indeed occur at (almost) every effective bunch crossing. As explained
above, eikonalization basically combines two (or more) events with one pair
of minijets each into one event with two (or more) pairs of minijets. This
has very little impact on the underlying event, since in both cases the
number of minijets contained in it will be approximately the same, as will
be the particle multiplicity, the total transverse energy, etc. All these
quantities are approximately proportional to the product of the
total cross section and the jet multiplicity per interaction, which should
to good approximation be described by the leading order calculation.

This simple argument is at least to some extent borne out by a full Monte
Carlo simulation \cite{63} of minijet events at \ppbar\
colliders\footnote{We mention in passing that in this analysis \ptmin\
was fixed from the total charged particle multiplicity, not from the cross
section; again values in the range from 1.5 to 2 GeV were found for
$\rs \leq 1$ TeV}; multiple interactions lead
to higher multiplicities, and thus to larger
underlying events. Moreover, it is known experimentally \cite{53} that
not only the total inelastic \ppbar\ cross section, but also the average
charged particle multiplicity \nav\ per unit of rapidity as well as the average
transverse momentum \ptav\ of charged particles increase with energy.
If one has $\gg 1$ event per effective bunch crossing, the total scalar
\pT\ in the underlying event, which should be a good measure of the
background problems caused by it, is approximately proportional to the product
$\sigma \cdot \nav \cdot \ptav$; this product grows much more rapidly with
energy than the total cross section alone does. We therefore believe that
eq.(\ref{e29}) provides in most cases a good figure of merit for the
background problems caused by the underlying event, even for values of \wgg\
where it no longer describes the true total \gaga\ cross section.

This is not true, however, if one expects the average number $\langle n
\rangle$ of two--photon events per effective bunch crossing to be close to
1. In that case it might be important to know what fraction of bunch
crossings, and thus annihilation events, will be entirely free of
two--photon events, as discussed above. This fraction is given by
$e^{-\langle n \rangle}$, which varies rapidly with $\langle n \rangle$ if
$\langle n \rangle \simeq 1$; e.g. $e^{-1/2} = 0.61$, while $e^{-2} =
0.14$. In such a situation it is probably advisable to plan for the worst,
or else to postpone a final decision until the total \gaga\ cross section
can be predicted more reliably.

\section*{5. Summary and Conclusions}
\setcounter{footnote}{0}
In this paper we have studied various two--photon reactions
leading to hadronic final states at future
linear \ee\ and \gaga\ colliders. The photon spectrum at these machines will
be quite different from the Weizs\"acker--Williams bremsstrahlung spectrum
familiar from \ee\ storage rings. In case of the \ee\ linacs, an important
new contribution to the photon flux comes from beamstrahlung. We saw in sec. 2
that the shape and normalization of the beamstrahlung spectrum depends quite
sensitively on the size and shape of the electron and positron bunches. Already
at \rs\ = 500 GeV, the beamstrahlung contribution to the total photon flux
can be anywhere between almost negligible (except for very small photon
energies) and clearly dominant (except for very hard photons); as a result,
photon fluxes at existing designs of 500 GeV colliders differ by as much as
a factor of 30. It is therefore very difficult to make definite statements
about two--photon reactions at such colliders which are valid for all
possible designs; instead, we tried to give an impression of the range of
cross sections or event rates one may expect.

In sec. 3 we studied several hard two--photon reactions. We showed in sec. 3a
that even without beamstrahlung a large majority of all hard events at future
\ee\ colliders will come from two--photon reactions (unless a new $Z'$ gauge
boson is found). In principle, one could discard many of these events already
at the trigger level, by requiring a large transverse or total energy in the
event. This might be dangerous, however, since ``new physics'' events
containing heavy stable neutral particles would also have visible energy
substantially below \rs. Moreover, two--photon events are interesting in their
own right. In particular, it is important to study the transition from
``hard'' to ``soft'' events; more about that later. The price one has to
pay for a low trigger threshold is that one has to deal with a very large
number of events; for instance, one expects at least 4 million events per
year with transverse energy $E_T \geq 10$ GeV even at the TESLA collider,
which has the smallest beamstrahlung of all designs we studied. Such event
rates are not unusual for hadron colliders, and pose no great
technological problems; such numbers might be unexpected for high energy
\ee\ colliders, however.  We described these events in some detail, giving
distributions of variables of interest (transverse momentum, rapidity and
di--jet invariant mass). In particular, we showed that at machines with
very strong beamstrahlung, exemplified here by the Palmer G design (see
Table 1), it would be difficult to study annihilation events containing a
hard photon collinear with the beam pipe and a real $Z$ boson.  Such
events are interesting \cite{38} since they would allow to study the QCD
evolution of hadronic systems between two quite different energy scales
($m_Z$ and 500 GeV, respectively) in one experiment; they might also allow
to self--calibrate the detector \cite{28}.

At future \ee\ linacs a large number of soft \ee\ pairs will be produced at
each bunch crossing. Most of these pairs will emerge at small angles. These
very large electron and positron fluxes probably force one to leave
substantial dead zones around the beam pipes when constructing detectors
for such colliders. One then expects a substantial cross section for
mono--jet events; these can be produced from di--jet final states
where one jet emerges at a small angle while the second
jet is produced centrally. Such one--sided events are a signature for
some ``new physics'' processes, like the production of supersymmetric
neutralinos; it is therefore important to know the standard model prediction
for this final state accurately. We found in sec. 3b that at \rs\ = 500 GeV,
it is dominated by annihilation events if $\pT > 30$ GeV; at higher energies,
however, two--photon events dominate over most of the kinematically accessible
region. Their cross section is well above that of typical new physics
processes. When looking for such processes one might therefore have to require
the missing \pT\ to be larger than the value that can be produced by boosted
two--jet events.

Two--photon events also produce a large majority of all \ccbar\ and \bbbar\
pairs at future \ee\ colliders. The total cross sections, listed in tables
2 and 3, are very large; e.g., even at the TESLA collider at least 80
million \ccbar\ and 400,000 \bbbar\ events would be produced per year.
However, without a full detector--specific Monte Carlo study it is impossible
to say what fraction of these events will be identified or even detected.
In contrast, \ttbar\ production will be dominated by annihilation events
at least up to \rs\ = 1 TeV, at all the proposed accelerators we studied.

Finally, two--photon processes can also contribute to the single production
of $W$ and $Z$ bosons. At \rs\ = 500 GeV, the rates are still marginal,
except for the case of the \gaga\ collider. Even at higher energy \ee\
colliders these process will not be able to compete with the single production
of $W$ and $Z$ bosons from $2 \rightarrow 3$ processes like $\ee \rightarrow
e W \nu$, as far as the total cross section is concerned. However, the
two--photon events always contain some hadronic activity, and produce
gauge bosons with small transverse momentum, in contrast to the $2 \rightarrow
3$ reactions. It is therefore important to include the two--photon processes
in a complete simulation of $W$ and $Z$ production.

In all cases we included direct as well as resolved photon contributions
when estimating two--photon cross sections. The relative importance of these
two classes of contributions depends on the process under consideration, on
the photon spectrum, as well as on the region of phase space one is studying.
If a given final state can be produced via gluon exchange in the $t$ or $u$
channel (e.g., jet production), the resolved photon contributions are more
important than for reactions that can only proceed via $s$ channel and
quark exchange diagrams (e.g., heavy quark production). Moreover, harder
photon spectra favour resolved photon events over direct ones, since at
higher photon energies one probes the parton densities inside the photon at
smaller values of Bjorken--$x$, where they increase rapidly. This can also
be achieved by going to particular regions of phase space, which favour
asymmetric initial state configurations; in this case, once resolved
contributions are very important. As a rule we find that at \ee\ colliders
operating at \rs\ = 500 GeV, resolved photon contributions never dominate
if the typical momentum scale of the process exceeds 40 to 50 GeV; in
heavy quark production they become subdominant already for $\pT \geq 10$ GeV.
Of course, there are also final states which in leading order cannot be
produced in direct two--photon reactions, like the vector quarkonium states
discussed in sec. 3c and the $W$ and $Z$ bosons of sec. 3d.

We also find that beamstrahlung can have a
significant effect on the annihilation
cross section for events with visible energy well below \rs; it can also
affect the total cross section for the production of a given final state,
as shown in the case of \ttbar\ production in sec. 3c. However, in all
cases beamstrahlung increases the two--photon cross section much more than
the annihilation cross section.

In sec. 4 we discussed semi--hard and soft two--photon reactions in some
detail. We showed that at certain designs one has to expect several such events
to occur simultaneously (within the time resolution of the detector) with any
annihilation event, giving rise to an ``underlying event''. Going to the
\gaga\ collider option only makes this problem worse. These qualitative
conclusions are independent of whether one estimates the total cross section
using semi--hard QCD (minijets), or relies on the VMD estimate. We presented
arguments showing that from the point of view of perturbative QCD, it is quite
natural to expect the \gaga\ cross section at high energies to substantially
exceed the VMD prediction, perhaps by as much as a factor of 10 or more.
Furthermore, once an underlying event occurs, quantities like the total
particle multiplicity and the total (transverse) energy in the underlying
event are of more immediate experimental interest than the number of separate
two--photon reactions that contributed to it. These quantities are
proportional to the product of the total \gaga\ cross section and the particle
multiplicity or (transverse) energy per interaction; in the case of \ppbar\
collisions these quantities are known experimentally to increase much more
rapidly with energy than the total cross section does. We therefore argued
that the simple leading order estimate for the cross section provides a
good figure of merit for the severity of the problems caused by the underlying
event, even if it does not reproduce the total \gaga\ cross section
accurately.

A hard beamstrahlung spectrum also has other adverse effects. As already
mentioned, beamstrahlung depends sensitively on the size and shape of the
bunches; it also depends on the bunch overlap during collisions. (The
expressions of sec. 2 always assume perfect overlap, i.e. fully central bunch
collisions.) The same factors also determine the luminosity. Therefore the
number of two--photon events due to beamstrahlung grows much more rapidly than
linearly with the luminosity per bunch crossing. One will then have to
accurately keep track of fluctuations and systematic changes of the
luminosity in order to estimate two--photon backgrounds and
signal--to--noise ratios precisely; this poses new challenges to the
construction of realistic event generators. Finally, as well known
\cite{1a,23}, beamstrahlung is also responsible for the large number of
soft \ee\ pairs mentioned above, which cause a multitude of technological
and physics problems; and by smearing out the electron beam energy, it
makes it difficult to study new thresholds in detail.

Of course, it has to be admitted that at present our predictions for hard as
well as semi--hard two--photon cross sections suffer from several
uncertainties.  In the case of hard resolved photon events the biggest
unknown is the parton content of the photon; in case of the gluon, it is
at present only known to at best a factor of 2. We saw in sec. 3a that
present parametrizations for
\Gg\ even differ by as much as a factor of 5 at low Bjorken--$x$ and small
momentum scale $Q^2$. This leads to large uncertainties in the predictions for
total jet rates for $\pT \leq 20$ GeV (for \rs\ = 500 GeV), as well as for
total \ccbar\ and \bbbar\ production rates. Fortunately, this situation should
improve soon. In the near future, studies of
heavy quark and jet production at TRISTAN and LEP \cite{6,63a} as well as
HERA \cite{64} will provide new information on the hadronic structure of
the photon. In a few years valuable new information should also come from
measurements of the photon structure function $F_2^{\gamma}$ at LEP \cite{65}
 in the
region of small $x$; in this region the evolution equations lead to a strong
coupling of quark and gluon densities, while $F_2^{\gamma}$ at large $x$
is not very sensitive to \Gg. Measurements of deep inelastic scattering
have the advantage that higher order corections (``k-factors'') are
expected to be smaller than in case of real \gaga\ scattering. The ultimate
$F_2^{\gamma}$ measurement might come \cite{66} from $e \gamma$ colliders,
which are a hybrid of the \ee\ and \gaga\ colliders discussed in this paper.

More experimental information about the details of the spectator jets from
resolved photons, as well as of the behaviour of total cross sections for
hadronic processes involving real photons in the initial state, is also
needed. TRISTAN can make important contributions also in this area, since
it can study semi--hard two--photon events at energies where eikonalization
does certainly not play a major role. This should allow to determine the
cut--off parameter \ptmin\ with greater confidence; we saw that already
the first AMY measurement \cite{7}, which is based on some 300 events, was very
helpful in this respect. A more detailed study of resolved photon events
might necessitate to upgrade the detectors towards a better angular coverage,
so that the spectator jets can be reconstructed more completely. In this area
HERA seems to have an advantage. According to first Monte Carlo studies
\cite{67}, HERA detectors should be able to isolate resolved photon events
efficiently and reliably; the large cross sections expected at HERA should
then allow detailed investigations of these events. Furthermore, as
already mentioned in sec. 4, at HERA the total $\gamma p$ cross section will
be measured at energies up to about 250 GeV; this should help to weed out
some of the existing ans\"atze \cite{60,61,62} for the eikonalization of
photon cross sections. In some models \cite{60,storrow} a first hint of
eikonalization might even be visible at LEP200.

We already argued in sec. 4 that the presence of an underlying event would
introduce many problems familiar from hadron colliders. In addition, it
would become impossible to distinguish between hard resolved and direct
two--photon reactions on an event by event basis; the spectator jets which
are the tell--tale signature for the former would be lost among the soft
hadrons of the underlying event. We therefore see that soft and semi--hard
two--photon reactions can commit some sort of fratricide by making the detailed
study of hard two--photon events very difficult. Even the measurement of the
total \gaga\ cross section, which is of great theoretical interest, is much
easier if the probability to have more than one event per bunch crossing is
very small, since only in this case the total cross section is directly
proportional to the number of bunch crossings that contain some hadronic
acitivity. Moreover, the ability to
trigger against the presence of a spectator jet would (greatly) reduce many
two--photon induced hard backgrounds; essentially one would only have to
deal with the direct contributions, the cross sections of which can be
calculated almost unambiguously. In fact, one can probably remove almost
the whole mono--jet background to true one--sided events if the presence
of a forward jet is detectable; we argued in sec. 3b that in principle such
a jet should be visible, {\em if} it is not totally obscured by an underlying
event.

In view of the undesirable consequences of having $\geq 1$ event per bunch
crossing, the most conservative attidute seems to be to design colliders
such that there is a large safety margin, i.e. {\em not} to rely on the
``conservative'' VMD model prediction, nor on calculations of eikonalized
cross sections that make use of it. Fortunately, we saw in sec. 4 that it seems
to be possible at least in principle to extend the superconducting TESLA
design to \rs\ = 2 TeV and beyond without risking the occurence of an
underlying event. This should not be misunderstood as our endorsement of
a particular design; rather it is an (at least theoretical) existence proof
for designs that maintain the traditional ``clean'' environment of \ee\
colliders up to TeV energies, as far as hadronic backgrounds are
concerned. Moreover, even an \ee\ collider where ${\cal O} (1)$ two--photon
event underlies every annihilation event has many advantages over hadron
colliders, because at \ee\ colliders the cross section for the production
of almost any heavy new particle will be at least roughly comparable to
the cross section for typical standard model annihilation processes; at
hadron colliders this is only true if the new particle carries colour.
Therefore we do {\em not} believe that soft and semi--hard two--photon
events will be the demise of the \ee\ collider; they do, however, provide
a strong additional argument in favour of designs with low beamstrahlung.

\subsection*{Acknowledgements}
We thank P. Zerwas and E. Levin for discussions on total cross sections,
K. Yokoya for helpful clarifications of some problems caused by
beamstrahlung as well as ongoing efforts to control them, and A. Djouadi for
discussions on three--jet production in \ee\ annihilation.

\noindent

\clearpage
{\bf Table 1:} Parameters of the machine designs we use. P--G, P--F, D--D and
T stand for Palmer G, Palmer F, DESY--Darmstadt and TESLA, respectively;
JLC1,2,3 stands for the three phases of the Japan Linear Collider. Notice that
there are two versions of the DESY--Darmstadt design, the original wide band
beam (wbb) design \cite{28} and its narrow band beam (nbb) variant \cite{DDP}.
\Y\ is the
beamstrahlung parameter, $\sigma_z$ the bunch length, \lhat\ the luminosity per
bunch crossing, N the number of bunches in each train, $\Delta t$ the temporal
separation between two conecutive bunches in one train, and ${\cal L}$ the
luminosity of the collider.
\begin{center}
\begin{tabular}{|c||c|c|c|c|c|c|c|c|}
\hline
  & P--G & P--F & D--D(wbb) & D--D(nbb) & T & JLC1 & JLC2 & JLC3\\
\hline
$\Upsilon$ & 0.42 & 0.11 & 0.065 & 0.015 & 0.0083 & 0.118 & 0.404 & 0.613\\
$\sigma_z \ [mm]$ & 0.11 & 0.11 & 0.4 & 1.0 & 2.0 & 0.152 & 0.113 & 0.095\\
$\tilde{{\cal L}} \ [\mu b^{-1}]$ & 5  & 1.1 & 0.29 & 0.12
& 0.26 & 0.8 & 2.9 & 4.2\\
N & 10 & 10 & 172 & 172 & 800 & 20 & 20 & 20\\
$\Delta t \ [10^{-9}sec]$ & 1 & 1 & 11 & 11 & 1000 & 1.4 & 1.4 & 1.4 \\
${\cal L} [10^{33}/ cm^2 sec]$ & 5.9 & 1.4 & 2.6 & 1.0
& 2.1 & 2.4 & 8.8 & 12.7 \\
\hline
\end{tabular}
\end{center}

\vspace*{6mm}
\noindent
{\bf Table 2:} Total \ccbar\ cross sections from two--photon processes at
7 \ee\ colliders of table 1, as well as for a ``\gaga'' collider made from an
\ee\ collider with \rs\ = 500 GeV; results for the DESY--Darmstadt (nbb)
design are very close to those for the TESLA design.
We have used the DG parametrization to
estimate the resolved photon contributions. $\sigma( \qqbar)$ and $\sigma(gg)$
stand for the 2--res \qqbar\ annihilation and gluon fusion cross sections,
$\sigma(\gamma g)$ for the 1--res photon gluon fusion cross section, and
$\sigma(\gaga)$ for the direct cross section; $\sigma(J/\psi)$ is the 1--res
$\gamma + g \rightarrow J/\psi + g$ cross section in the color singlet model.
All cross sections are in nb.
\begin{center}
\begin{tabular}{|c||c|c|c|c|c|c|}
\hline
Collider & $\sigma(\qqbar)$ & $\sigma(gg)$ & $\sigma(\gamma g)$ &
$\sigma(\gaga)$ & $\sigma($tot) & $\sigma(J/\psi)$ \\
\hline
T & 0.010 & 0.038 & 1.8 & 2.2 & 4.0 & 0.014 \\
D--D(wbb) & 0.041 & 0.11 & 7.0 & 6.4 & 13.5 & 0.053 \\
P--F & 0.017 & 0.08 & 4.0 & 2.4 & 6.4 & 0.030 \\
P--G & 0.14 & 1.1 & 38 & 9.9 & 49 & 0.28 \\
JLC1 & 0.029 & 0.12 & 6.3 & 3.7 & 10.1 & 0.047 \\
JLC2 & 0.064 & 1.3 & 31 & 3.9 & 36 & 0.22 \\
JLC3 & 0.054 & 2.2 & 41 & 3.1 & 46 & 0.28 \\
\gaga(500) & 0.13 & 7.6 & 130 & 0.14 & 140 & 0.89 \\
\hline
\end{tabular}
\end{center}

\clearpage
\vspace*{6mm}
\noindent{\bf Table 3:} Total cross sections for \bbbar\ production from
two--photon processes. The notation is as in table 2, except that now all
cross sections are in pb.
\begin{center}
\begin{tabular}{|c||c|c|c|c|c|c|}
\hline
Collider & $\sigma(\qqbar)$ & $\sigma(gg)$ & $\sigma(\gamma g)$ &
$\sigma(\gaga)$ & $\sigma($tot) & $\sigma($\Y(1s))\\
\hline
T & 0.39 & 0.46 & 10.4 & 7.6 & 19 & 0.026 \\
D--D(wbb) & 2.0 & 1.2 & 38 & 30 & 71 & 0.097 \\
P--F & 1.1 & 1.1 & 26 & 12 & 41 & 0.066 \\
P--G & 10 & 13 & 260 & 65 & 350 & 0.66 \\
JLC1 & 1.8 & 1.5 & 39 & 20 & 62 & 0.10 \\
JLC2 & 6.4 & 24 & 280 & 26 & 330 & 0.66 \\
JLC3 & 7.3 & 51 & 430 & 20 & 510 & 1.0 \\
\gaga(500) & 21 & 150 & 1,300 & 4.2 & 1,500 & 3.5 \\
\hline
\end{tabular}
\end{center}

\noindent
{\bf Table 4:} Total cross sections for single production of $W$ and $Z$
bosons in \gaga\ collisions at various colliders,
estimated from the twice resolved $\qqbar
\rightarrow W, Z$ contribution. The $W$ cross section inlcudes both $W^+$
and $W^-$ production. We have used the DG parametrization with
$Q^2 = m^2_{W,Z}$. All cross sections are in fb.
\begin{center}
\begin{tabular}{|c||c|c|}
\hline
Collider & $\sigma(W)$ & $\sigma(Z)$ \\
\hline
T & 2.0 & 1.0 \\
D--D (wbb) & 5.0 & 2.3 \\
P--F & 4.7 & 2.2 \\
P--G & 58 & 28 \\
JLC1 & 6.5 & 3.0 \\
JLC2 & 77 & 39 \\
JLC3 & 115 & 55 \\
\gaga(500) & 400 & 205 \\
\gaga(1000) & 800 & 340 \\
\gaga(2000) & 1,750 & 615 \\
\hline
\end{tabular}
\end{center}
\clearpage
\noindent
{\bf Table 5:} Total semi--hard two--photon cross section
at various colliders. The notation for the first 9 rows is like in the
previous tables; rows 10 and 11 show results for an upgrade of the TESLA
design where the beamstrahlung parameter \Y\ grows like \rs, while the
last two rows are for an upgraded TESLA if \Y\ grows like $s$.
We have chosen $\ptmin = 1.6 \ (2.4)$
GeV for the DG (DO+VMD) parametrization, as described in the text. For
comparison, col. 4 shows the soft contribution for $\wgg \geq 5$ GeV, assuming
a constant \gaga\ cross section of 250 nb as predicted by the VMD model.
Col. 5 shows the number of
semi-hard events per bunch collision or per $10^{-7}$ sec, whatever is bigger.
\begin{center}
\begin{tabular}{|c||c|c|c|c|}
\hline
 Collider & $\sigma^{hard}({\rm DG}) \ [\mu b]$ &
$\sigma^{hard}({\rm DO+VMD}) \ [\mu b]$ & $\sigma^{soft} \ [\mu b]$ &
no. of events (DG) \\
\hline
T & 0.016 & 0.0090 & 0.041 & 0.004 \\
D--D (nbb) & 0.020 & 0.014 & 0.051 & 0.021 \\
D--D (wbb) & 0.075 & 0.041 & 0.20 & 0.20 \\
P--F & 0.042 & 0.024 & 0.072 & 0.46 \\
P--G & 0.48 & 0.29 & 0.51 & 24 \\
JLC1 & 0.069 & 0.04 & 0.12 & 1.1 \\
JLC2 & 0.41 & 0.28 & 0.19 & 24 \\
JLC3 & 0.59 & 0.43 & 0.15 & 50 \\
\gaga(500) & 1.9 & 1.4 & 0.25 & 0.49 -- 95 \\
T(1000) & 0.057 & 0.036 & 0.099 & 0.0036 \\
T(2000) & 0.21 & 0.15 & 0.13 & 0.013 \\
T'(1000) & 0.17 & 0.099 & 0.27 & 0.043 \\
T'(2000) & 3.4 & 2.4 & 1.2 & 3.4 \\
\hline
\end{tabular}
\end{center}
\newpage
\section*{Figure Captions}

\renewcommand{\labelenumi}{Fig. \arabic{enumi}}
\begin{enumerate}
\item  %Fig.1
Photon spectra at \rs\ = 500 GeV (a) and evolution of the photon spectrum with
\rs\ at the JLC design (b). T, D--D, P--F and P--G stand for the TESLA,
DESY--Darmstadt, Palmer F and Palmer G designs, respectively; note that the
DESY--Darmstadt design exists in wide band beam (wbb) and narrow band beam
(nbb) versions. WW is the
Weizs\"acker Williams or bremsstrahlung spectrum. The curve labelled `laser'
shows the spectrum (\ref{e8}) that emerges when laser photons are backscattered
off incident electrons.

\vspace*{6mm}
\item   %Fig.2
Electron spectra at \rs\ = 500 GeV. The dotted curve shows the electron
spectrum without beamstrahlung, but with initial state radiation included.
The notation for the other curves is as in fig. 1. Notice that we have
chosen to present the spectra as a function of $1-x$, in order to better
resolve the region of large $x$.

\vspace*{6mm}
\item   %Fig.3
Cross sections for the two--photon production of two central jets at \rs\ =
500 GeV (a) and the three stages of the JLC (b), as a function of the minimal
transverse momentum \ptmin\ of the jets. The notation in (a) is like in
Fig. 1a;
notice that the results for the DESY--Darmstadt (nbb) design are almost
identical to those for the TESLA collider.
In (b), the dashed curves show the prediction from the direct process alone,
while the solid curves show the prediction after inclusion of the
resolved photon contributions. The DG parametrization has been used with
$Q^2 = \hat{s}/4$ and a floating number of flavours, as described in the text.

\vspace*{6mm}
\item   %Fig. 4
Various contributions to the two--photon production of two central jets at
the first stage of the JLC. In (a) only the once resolved contributions are
shown, while (b) depicts the twice resolved contributions. The curves are
labelled according to the composition of the final state; here `q' stands
for any quark or anti--quark. We have used the same parameters as in fig. 3.
In particular, the use of a $Q^2$ dependent number of flavors explains the
kinks at $\pT \simeq 7$ GeV, where charm starts to contribute.

\vspace*{6mm}
\item     %Fig.5
Ratios of predictions of the LAC2 and DG parametrizations for the production
of two central jets at the first stage of the JLC. Contributions with
different final states have been shown separately, using the same notation as
in fig. 4.

\vspace*{6mm}
\item    %Fig. 6
Rapidity distribution of di--jet events at \pT\ = 30 GeV for the case $y_1 =
y_2$, at the TESLA collider
(a) and a 500 GeV \gaga\ collider (b). The contributions from the direct, once
resolved and twice resolved processes are shown separately. We have used the
same parameters as in fig. 3. Notice that (b) includes a very small, but
nonzero
direct contribution (short dashed curve).

\vspace*{6mm}
\item    %Fig. 7
Invariant mass distribution of centrally produced di--jet events at the Palmer
G (a) and TESLA (b) colliders. The solid and dashed curves show the two--photon
contributions in the notation of fig. 6, while the dotted curves show the
contribution from annihilation events. Notice that a stronger \pT\ cut
has been applied in (a). We have used the same parameters as in fig. 3.

\vspace*{6mm}
\item     %Fig. 8
The transverse momentum spectrum of mono--jets, at two different \ee\ colliders
operating at \rs\ = 500 GeV (a), as well as a \gaga\ collider (b). The solid
and dotted curves in (a) show the total two--photon and annihilation
contributions, respectively, while the various curves in (b) correspond to
different classes of two--photon contributions; in (b) there is no annihilation
contribution, of course. We have used the same parameters as in fig. 3.

\vspace*{6mm}
\item   %Fig. 9
The transverse momentum spectrum of mono--jets at the three stages of the
JLC. Notations and parameters are as in fig. 8a.

\vspace*{6mm}
\item     %Fig.10
The transverse mometum spectrum of central charm pairs produced from
two--photon processes at the TESLA collider (a) and at Palmer G (b),
respectively. Contributions from different classes of processes are
shown separately; notice that in this figure, the two twice resolved
contributions are labelled according to the initial state. We have used the
DG parametrization with $N_f$ = 3 active flavours.

\vspace*{6mm}
\item    %Fig. 11
Total \ttbar\ production cross sections at two different \ee\ colliders
operating at \rs\ = 500 GeV (a) and at the third stage of the JLC (b). The
dotted and solid curves in (a) show contributions from the annihilation
process and from two--photon reactions, respectively; notice that the latter
have been multiplied with 100 (10) for the TESLA (Palmer G) collider. In (b)
we show in addition the annihilation contribution if both initial state
radiation and beamstrahlung could be switched off (long dashed curve), as well
as the contribution from the direct two--photon process (short dashed curve).

\vspace*{6mm}
\item     %Fig. 12
Total integrated semi--hard (minijet) two--photon cross section as a function
of the transverse momentum cut--off parameter \ptmin, for the DG (a) and
modified DO+VMD (b) parametrizations. The notation is the same as in fig. 1a.

\vspace*{6mm}
\item    %Fig. 13
The total semi--hard \gaga\ cross section as predicted by the DG
parametrization with \ptmin\ = 1.6 GeV, as a function of the \gaga\
centre--of--mass energy \wgg\ (solid). The dashed curves show which
fraction of all semi--hard two--photon events at a given collider have a
\gaga\ energy less than \wgg; they refer to the scale at the right.

\end{enumerate}
\end{document}